\documentclass[12pt,a4paper]{article}
\usepackage{subcaption}

\usepackage{braket} 
\usepackage{caption}
\usepackage{jheppub}
\usepackage{float}
\usepackage[utf8]{inputenc}
\usepackage{csquotes}
\DeclareUnicodeCharacter{2212}{\textendash}

\title{\boldmath{Spread complexity for measurement-induced non-unitary dynamics and Zeno effect}
}
 
\author[a,b]{Aranya Bhattacharya,}
\author[c]{Rathindra Nath Das,}
\author[d]{Bidyut Dey,}
\author[c]{and \\Johanna Erdmenger}

\affiliation[a]{Institute of Physics, Jagiellonian University, Łojasiewicza 11, 30-348 Kraków, Poland}
\affiliation[b]{Centre for High Energy Physics, Indian Institute of Science,\\ C.V. Raman Avenue, Bangalore 560012, India.}
\affiliation[c]{Institute for Theoretical Physics and Astrophysics and \\ W\"urzburg-Dresden Cluster of Excellence ct.qmat \\Julius-Maximilians-Universit\"at W\"urzburg \\
 Am Hubland, 97074 W\"{u}rzburg, Germany}
\affiliation[d]{Indian Institute of Technology - Kanpur, Kanpur 208016, Uttar Pradesh, India}

\emailAdd{aranya.bhattacharya@th.if.uj.edu.pl, \\das.rathindranath@uni-wuerzburg.de,\\ bidyutd@iitk.ac.in,\\erdmenger@physik.uni-wuerzburg.de }

\abstract{Using spread complexity and spread entropy, we study non-unitary quantum dynamics.  For non-hermitian Hamiltonians, we extend the bi-Lanczos construction for the Krylov basis to the Schrödinger picture. Moreover, we implement an algorithm adapted to complex symmetric Hamiltonians. This reduces the computational memory requirements by half compared to the bi-Lanczos construction. We apply this construction to the one-dimensional tight-binding Hamiltonian
subject to repeated measurements at fixed small time intervals, resulting in effective
non-unitary dynamics. We find that the spread complexity initially grows with time, followed by an extended decay period and saturation. The choice of initial state determines the saturation value of complexity and entropy. In analogy to measurement-induced phase transitions, we consider a quench between hermitian and non-hermitian Hamiltonian evolution induced by turning on regular measurements at different frequencies.  We find that as a function of the measurement frequency, the time at which the spread complexity starts growing increases.  This time asymptotes to infinity when the time gap between measurements is taken to zero, indicating the onset of the quantum Zeno effect, according to which measurements impede time evolution. 
}

\begin{document}
\maketitle
\flushbottom

\section{Introduction}
The application of information-theoretic tools to quantum systems is one of the major current research  directions in theoretical physics, ranging from simple models in quantum mechanics to complicated quantum field theoretic and holographic setups. One of the recent advances in this context is the study of complexity, in order to characterise the evolution of a quantum state or operator. Complexity is of particular interest in the context of holographic conjectures \cite{Susskind:2014rva, Brown_2016}. Motivated by holography, Nielsen's circuit complexity \cite{Nielsen_2006}, which is the first example relating complexity to geometric concepts such as geodesics, was investigated for new quantum mechanical and quantum field theoretical systems \cite{Jefferson:2017sdb, PhysRevLett.120.121602, PhysRevLett.122.231302, Flory:2020dja, Erdmenger:2020sup, Bhattacharya:2022wlp}\footnote{For detailed review on these developments, see the review \cite{Chapman:2021jbh} and references therein.}.  An alternative proposal of complexity, free of  ambiguities such as the choice of gates or reference state, is Krylov complexity \cite{Parker:2018yvk}. Krylov complexity is defined for an orthonormal basis of operators (states) in the Heisenberg (Schrödinger) picture using the Lanczos algorithm \cite{VM,Parker:2018yvk, Balasubramanian:2022tpr} which quantifies the growth of an operator (state) during time evolution. While most of the studies for Krylov complexity to date have been for operators \cite{Barbon:2019wsy, Dymarsky:2019elm, Dymarsky:2021bjq, Yates:2021asz, Caputa:2021sib, Kim:2021okd, Caputa:2021ori, Bhattacharjee:2022vlt, Bhattacharjee:2022ave, Patramanis:2021lkx, Trigueros:2021rwj, Rabinovici:2021qqt, Fan:2022xaa, Heveling:2022hth, Bhattacharyya:2023dhp, Muck:2022xfc, Kundu:2023hbk, Banerjee:2022ime, Du:2022ocp, Guo:2022hui, Avdoshkin:2022xuw, Bhattacharyya:2020qtd, Bhattacharya:2023xjx, Iizuka:2023pov, Iizuka:2023fba, Camargo:2023eev, Camargo:2022rnt}, in \cite{Balasubramanian:2022tpr}  an equivalent definition for Schrödinger evolution of quantum states was proposed using the Krylov basis. This is known as the Krylov spread complexity, which quantifies how the information of a quantum state spreads under Hamiltonian evolution. This spreading implies that the state becomes more complex. However, as the physical Hamiltonian is usually hermitian, these studies considered unitary evolution of quantum states  \cite{Caputa:2022eye, Afrasiar_2023, Balasubramanian:2022dnj, Caputa:2022yju, Erdmenger:2023shk, Pal:2023yik, Nandy:2023brt, Chattopadhyay:2023fob, Gautam:2023bcm, Bhattacharjee:2022qjw, Gill:2023umm, Bento:2023bjn, Aguilar-Gutierrez:2023nyk, Craps:2023ivc}. 

In this paper, we address the question of how the Krylov spread complexity behaves for non-unitary dynamics. In principle, the methods we develop are general and can be applied to any non-unitary evolution of quantum state. For definiteness, we focus on a specific case known as the quantum first passage problem (QFPP), where the dynamics become non-unitary due to repeated projective measurements \cite{Peres, Dhar2013QuantumTO, Dhar_2015, Anastopoulos_2006, Erez_2008, PhysRevE.95.032141}. It is a well-known fact from the postulates of quantum mechanics that these projective measurements are non-unitary operations, and hence, the evolution of the system becomes non-unitary. The effect of the measurement is captured perturbatively by an effective non-hermitian Hamiltonian, where the time between two consecutive measurements acts as the perturbation parameter.

Previously, Krylov complexity was investigated for unitary evolution of isolated systems \cite{Parker:2018yvk, Rabinovici:2021qqt}, as well as for open system non-unitary dynamics of operators \cite{Bhattacharya:2022gbz, liu2022krylov, Bhattacharjee:2022lzy, Bhattacharya:2023zqt, Bhattacharjee:2023uwx}. Moreover, an approach towards non-unitary dynamics for operator growth of open quantum systems was given in \cite{Bhattacharjee:2022lzy, Bhattacharya:2022gbz, Bhattacharya:2023zqt}. These works are concerned with the operator Hilbert space, where the evolution is generated by the Liouvillian superoperator $\mathcal{L}=[H,.]$ instead of the Hamiltonian $H$. 

The starting point of the present work is the tight-binding Hamiltonian \cite{Dhar2013QuantumTO, Dhar_2015}. Subject to the measurements of the QFPP\footnote{The quantum first passage problem (QFPP) considers a quantum state evolving under a lattice Hamiltonian subject to local projective measurements performed on particular lattice sites \cite{Dhar_2015}.}, when the perturbative treatment is applied for modelling the non-unitary evolution, we obtain a non-hermitian Hamiltonian that we use to evolve position eigenstates. We obtain the non-hermitian total probability, the spread complexity, and the spread entropy to identify the characteristics of non-unitarity by comparing them to a scenario of unitary evolution without measurements.

We build upon the methodology of the bi-Lanczos algorithm previously utilized in \cite{Bhattacharya:2023zqt} for the non-unitary evolution of operator complexity. We extend this approach to investigating the state complexity for the non-unitary evolution of a quantum system subjected to projective measurements at regular intervals. Our analysis pursues a dual objective; first, to extend the non-unitary bi-Lanczos approach to spread complexity, and second, to explore the  behaviour of spread complexity for measurement-induced quantum channels. Additionally, we demonstrate how a specialized algorithm designed for complex symmetric matrices can effectively halve the workload and memory requirements. We refer to this new algorithm as a complex symmetric Lanczos algorithm. Applying these techniques to the QFPP, we analyze the spread complexity and entropy in the Krylov basis. Our investigation encompasses both open and periodic boundary conditions, revealing that the behaviour of spread complexity in this non-hermitian system is marked by an initial growth followed by an extended decay region and saturation. Notably, the prolonged decay region, as compared to unitary evolution, represents a novel complexity behaviour induced by measurements in this non-hermitian context. We also observe that different choices of the initial state yield distinct dynamics and saturation values for complexity and entropy. 

Finally, we explore a quench scenario, gradually changing the non-hermitian perturbation parameter from zero to higher values, effectively increasing the time gap between measurements in the system. By selecting eigenstates of the tight-binding hermitian Hamiltonian as the initial states, we examine the evolution of spread complexity under the effective non-hermitian quenched Hamiltonian. In this scenario, for the single-particle ground state, the  complexity dynamics shifts from oscillatory behaviour to an initial phase of rapid growth, followed by decay and saturation. For the single-particle first excited state as the initial state, the growth of the spread complexity is delayed when the time between two measurements is decreased. This indicates  a transition in the complexity behaviour based on the change in the measurement frequency. This is reminiscent of a measurement-induced phase transition \cite{measure1, measure2,Li_2018,Li_2019,Sang_2021}. For a very high frequency of measurements, we find that the initial state does not evolve in the Hilbert space. This is the characteristic of the quantum Zeno effect \cite{Sudarshan_1977,FACCHI200012,Facchi_2008}.


   Our effective non-hermitian Hamiltonian is a complex symmetric matrix. In addition to applying  the bi-Lanczos algorithm of \cite{Gaaf, 200SIAM, Gruning2011ImplementationAT}, which builds upon the open system operator complexity, we also implement our new complex symmetric Lanczos algorithm. We find that for complex symmetric Hamiltonians, the complex symmetric Lanczos works much more efficiently than the bi-Lanczos algorithm. This is due to the fact that for the bi-Lanczos algorithm, the computational power needed is larger due to the separate treatment of bra and ket states. Furthermore, unlike bi-Lanczos, the complex symmetric algorithm requires to deal with a single Krylov space only, and no conjugate Krylov space is needed.  In all the cases studied in this paper, we find exact numerical agreement between the results generated by the bi-Lanczos and the complex symmetric Lanczos methods. 

The paper is structured in the following way. We begin with a brief review of the Krylov spread complexity in section \ref{sec:Krylov-review}. In Section \ref{sec:Non-unitary algorith}, we discuss the modified non-hermitian norm in Krylov basis that we use for the non-unitary dynamics. We also review the bi-Lanczos algorithm and present the complex symmetric Lanczos algorithm. In section \ref{sec:setup_and_results}, we begin by reviewing the first passage problem of quantum mechanics in detail (section \ref{sec:first-passage}), followed by the main results of this work. These are divided into three parts. Section \ref{obc} contains our results for the one-dimensional chain with open boundary conditions for which we study the dependence of spread complexity and entropy on the non-hermiticity parameter. We also obtain the dependence of complexity and entropy on the spread of the initial state on the position eigenstate and the distance of the initial spread from the position of the detector. In section \ref{pbc}, we perform a similar study for the chain defined with periodic boundary conditions. For the periodic boundary conditions, we find cases where the evolution becomes similar to unitary evolution after some initial time window. In section \ref{quenched}, we consider two different eigenstates of the hermitian Hamiltonian before measurements as our initial states. We obtain the effects of a non-hermitian quench on the spread complexity under the influence of the measurement process.   Finally, we provide a detailed summary of results with future directions in section \ref{sec:conclusion}. In Appendix \ref{app:K_probability}, we find the time dependence of the time-dependent total probability and an analytical form for the return amplitude. In Appendix \ref{appb1}, we summarize the behaviour of the Lanczos coefficients, spread complexity and entropy for the unitary  tight-binding Hamiltonian. In Appendix \ref{appb2}, we discuss the form of the Lanczos coefficients for different non-hermitian cases.

\section{A brief review of spread complexity in the Krylov basis}\label{sec:Krylov-review}
The spread complexity in the Krylov basis was proven to be the most optimized measure of complexity that quantifies the spread of a quantum state as it evolves \cite{Balasubramanian:2022tpr}. The starting point is the Schrödinger evolution of a quantum state,
\begin{equation}\label{Schrodinger}
    |\psi(t)\rangle= e^{-i H t} |\psi(0)\rangle. 
\end{equation}
Below we briefly review the Schrödinger evolution in the Krylov basis \cite{Balasubramanian:2022tpr, Caputa:2022eye} and discuss the basic constituents needed to define the spread complexity in this basis. This is particularly useful for comparing with our new algorithms and normalisation for non-unitary evolution.

We note that any eigenstate of the Hamiltonian is not a good starting point for computing the spread complexity since the eigenstates of the Hamiltonian do not change under the Hamiltonian evolution apart from a phase proportional to the eigenvalue. According to \cite{Balasubramanian:2022tpr}, an appropriate starting point is to expand the exponential in Eq.~\eqref{Schrodinger},
     \begin{equation}
         |\psi(t)\rangle= |\psi(0)\rangle -i t H|\psi(0)\rangle + \frac{(it)^2}{2} H^2 |\psi(0)\rangle+\cdots +\frac{(i t)^n}{n!} H^n |\psi(0)\rangle+\cdots. \, , 
     \end{equation}
where the initial state is identified as the first element of the Krylov basis, $|K_0\rangle=|\psi(0)\rangle$. The further elements $|K_n\rangle$  are recursively constructed by implementing the generalized Lanczos algorithm with the Hamiltonian acting on any general element $|K_m\rangle$ of the Krylov space,
     \begin{equation}
         H|K_m\rangle=a_m|K_m\rangle +b_m |K_{m-1}\rangle+b_{m+1}|K_{m+1}\rangle \, ,
     \end{equation}
     where $a_m$ and $b_m$ are the Lanczos coefficients defined by
     \begin{align}
         \nonumber a_m &=\langle K_m|H|K_m\rangle, \, b_m= \langle A_m|A_m\rangle^{\frac{1}{2}}, \,\\ \text{with} ~~ |A_m\rangle &=(H-a_{m-1})|K_{m-1}\rangle -b_{m-1}|K_{m-2}\rangle.
     \end{align}
      For a hermitian Hamiltonian, the overall evolution is unitary, and the Lanczos coefficients $a_n$ and $b_n$ are both real. In terms of the constructed Krylov basis, the evolved state can be written in terms of the Lanczos basis vectors as
     \begin{equation}
         |\psi(t)\rangle= \sum_{n} \psi_n(t) |K_n\rangle. \label{k_com_unitary}
     \end{equation}
      These $|\psi_n(t)|^2$'s correspond to the probabilities of the state being in the $n$-th Krylov basis element $|K_n\rangle$ at time $t$ with the total probability being $P=\sum_{n} |\psi_n(t)|^2=1$.\footnote{This notion of Krylov probability shows decaying behaviour for non-hermitian cases, as shown in Appendix \ref{app:K_probability}, unless renormalized to one at all times.} The corresponding spread complexity is defined as the average position of the state in the Krylov space at time $t$. This, in the Schrödinger Krylov basis, is written as 
     \begin{equation}
         C_S(t)=\sum_{n} n|\psi_n(t)|^2.
     \end{equation}
      An alternative way to derive the Lanczos coefficients ($a_n, \, b_n$) is to start from the \textit{survival amplitude} $S(t)=\langle \psi(t)|\psi(0)\rangle=\psi_0(t)$ that measures the overlap between the initial and final states. Then the moments of the survival amplitude are calculated recursively method from the Hankel matrix formed out of  \cite{Parker:2018yvk, Balasubramanian:2022tpr}. The moments are defined by 
     \begin{equation}
         \mu_n=\frac{d^n}{dt^n} S(t)|_{t=0}.
     \end{equation}
     This moment recursion process can be in fact represented by an unnormalized Markov chain where the transition weights in different levels of the chain are simply the Lanczos coefficients ($a_n$ are the weights in the $n$th level and $b_n$ are the weights connecting $n$th level to the $(n+1)$-st one) \cite{Balasubramanian:2022tpr}.
     
      In analogy to the Krylov entropy for operator growth \cite{Barbon:2019wsy}, the spread entropy in the Schrödinger picture is given by
\begin{equation}
S_S(t) = -\sum_n |\psi_n(t)|^2 \ln |\psi_n(t)|^2.
\end{equation}
In fact, the spread complexity can be understood as an exponentiation of the spread entropy \cite{Balasubramanian:2022tpr}.

For chaotic evolution, the spread complexity shows linear growth up to exponential times concerning the degrees of freedom of the system, followed by saturation to a plateau after a brief decay phase \cite{Balasubramanian:2022tpr}. The corresponding Krylov spread entropy demonstrates logarithmic growth, followed by a region of linear growth before reaching a plateau for chaotic evolution. Furthermore, these Krylov spread measures, including probability, complexity, and entropy, can be utilized to characterize different quantum phases of matter \cite{Afrasiar_2023}, such as topological phases \cite{Caputa:2022eye}. Therefore, these measures in the Krylov spread basis are a good probe of various physical phenomena in unitarily evolving quantum systems.

\section{Non-unitary evolution and modified Lanczos algorithms\label{sec:Non-unitary algorith}} 
Here, we generalise concepts for the Krylov spread measures to the case of non-unitary evolution. We begin by presenting a modified normalisation for the density matrix in Krylov space. We continue by presenting two distinct algorithms for evaluating the Lanczos coefficients in the non-unitary case.

The unitarity of quantum mechanical  evolution arises from the hermiticity of the Hamiltonian $H$. Consequently, the evolution operator $U=e^{-i H t}$ is inherently unitary by definition. Nevertheless, instances exist where the effective evolution of a system deviates from unitarity. Two notable scenarios include i) open system dynamics: this pertains to the time dynamics of a subsystem within the broader system, exhibiting non-unitarity due to interactions with the larger universe and ii) measurements: non-unitarity emerges when measuring the state at any given moment.

Examining the characteristics of non-unitary evolution, we first consider the case of an effective non-hermitian Hamiltonian $H$ (with $H\neq H^{\dagger
}$), leading to a complex eigenspectrum expressed as $H=H_1+i H_2$, where both $H_1$ and $H_2$ are hermitian Hamiltonians. With this effective Hamiltonian, the evolution takes the form
\begin{equation}\label{changeofnorm}
|\psi(t)\rangle_{nh}= e^{-iHt}|\psi(0)\rangle_{nh}=e^{H_2t-iH_1t}|\psi(0)\rangle_{nh}.
\end{equation}
The non-hermitian part of the Hamiltonian, $H_2$, introduces a change in the normalisation of the state. 

\subsection{Revised normalisation and definitions of probability, complexity and entropy}\label{nhnewnorm}
According to \eqref{changeofnorm}, it is critical to adjust the normalisation continuously. To do so, we introduce a revised normalisation in the Krylov basis that renormalizes the state to unity at all times. The dynamical evolution of a non-hermitian system initialized in a density matrix $\rho_{nh}(0)=|\psi(0)\rangle_{nh}\langle \psi(0)|_{nh}$ is governed by \cite{newnorm}
\begin{equation}\label{nhdensity}
    \rho_{nh}(t)=\frac{e^{-iHt}\rho_{nh}(0)e^{iH^{\dagger}t}}{\text{Tr}[e^{-iHt}\rho_{nh}(0)e^{iH^{\dagger}t}]}.
\end{equation}
This ensures that the normalisation of the states under non-unitary evolution is one at any given moment in time. We implement this normalisation \eqref{newnorm} in the Krylov basis in the following way. We have the initial matrix wave function in the Krylov basis as 
\begin{equation}
    |\Phi_{nh}(0)\rangle_K= \begin{pmatrix} 1\\0\\0\\0\\\vdots\\0\\\end{pmatrix}_K,
    \label{eq:initialstateKry}\end{equation}
    which is equivalent to $\psi_n(t=0)=\delta_{n,0}$ for $\psi_n$ as given in  \eqref{k_com_unitary}, by identifying the first element of the column matrix as $\psi_0 (0)$. Now, accordingly, we can define a density matrix in Krylov space,
    \begin{equation}\label{norm_in_kry}
        (\rho_{nh})_K(0)=|\Phi_{nh}(0)\rangle_K\langle\Phi_{nh}(0)|_K=\begin{pmatrix} 1 &0&&&&0\\0&  0& 0&&&\\&0&0&\ddots &&\\&&\ddots &\ddots &0&\\&&&0&0&
    \ddots\\0&&&&\ddots&0\\\end{pmatrix}_{K\times K}.
    \end{equation}
    This Krylov space density matrix undergoes evolution with respect to the non-hermitian tridiagonalized form ($L$) of the non-hermitian Hamiltonian ($H$) that is derived using the appropriate Lanczos algorithm. This evolution is exactly similar to Eq.~\eqref{nhdensity}, with $L$ replacing $H$. Keeping the probability conserved at all times, we work with a revised time-dependent normalisation of a state vector in the Krylov basis and define the new state vector $|\widetilde{\Phi}(t)\rangle_K$ with probability $P(t)$ as follows,
\begin{equation}\label{newnorm}
    |\Tilde{\Phi}(t)\rangle_K= \frac{e^{-iLt} |\Phi_{nh}(0)\rangle_K}{\sqrt{\text{Tr}[e^{-iLt}(\rho_{nh})_K(0)e^{iL^{\dagger}t}]}}=\begin{pmatrix} \Tilde{\phi}_0(t)\\\Tilde{\phi}_1(t)\\\Tilde{\phi}_2(t)\\\Tilde{\phi}_3(t)\\\vdots\\\Tilde{\phi}_{K-1}(t)\\\end{pmatrix}_K,\, \,  P(t)=|_K\langle \Tilde{\Phi}(t)| \Tilde{\Phi}(t)\rangle_K|=1.
    \end{equation}
    We use this Krylov probability in our further definitions of spread entropy and complexity\footnote{Note that the square root over the trace in the denominator of Eq.~\eqref{newnorm} is unity when the revised normalisation is considered. Therefore this normalisation becomes exactly similar to the denominator of Eq.~\eqref{nhdensity}.}, 
    \begin{equation}\label{entcomp}
        C_S(t)= \frac{\sum_n n |\Tilde{\phi}_n(t)|^2}{P(t)}, \, \, S_S(t)= -\frac{\sum_n |\Tilde{\phi}_n(t)|^2 \ln |\Tilde{\phi}_n(t)|^2}{P(t)},
    \end{equation}
where $\Tilde{\phi}_n(t)$ is the $(n+1)$-th element of the matrix $|\Tilde{\Phi}(t)\rangle_K$. In terms of the normalisation given in Eq.~\eqref{newnorm}, the dynamically normalised probability $P(t)$ remains constant in time. Hence, the definitions in Eq.~\eqref{entcomp} become formally similar to the definitions for hermitian quantum systems as the denominator becomes $1$.


\subsection{Construction algorithms for the Krylov basis of non-hermitian systems}\label{sec:algorithms}
 As mentioned in section \ref{sec:Krylov-review}, one of the main features of the Krylov basis for hermitian Hamiltonians is that the Hamiltonian becomes tri-diagonal. We now summarize the details of the modifications on the Lanczos algorithm needed for the non-hermitian Hamiltonian. For non-hermitian matrices, the Arnoldi recursive technique \cite{Bhattacharya:2022gbz} is the simplest extension of the Lanczos algorithm. In this approach, the non-hermitian Hamiltonian becomes an upper Hessenberg matrix and thus not of tri-diagonal form. These extra terms impede the standard computation of Krylov complexity \cite{Bhattacharya:2023zqt}. 

\subsubsection{Bi-Lanczos algorithm}\label{sec:bilanczos}
This issue is resolved by the {\it bi-Lanczos algorithm} that we now implement for state complexity. The standard bi-Lanczos algorithm allows to tri-diagonalize a non-hermitian matrix $A$ \cite{Gaaf, 200SIAM, Gruning2011ImplementationAT}. Since a non-hermitian matrix $A$ acts differently on a ket vector $|v\rangle$ than on a bra vector $\langle w|$, finding an orthogonal basis set that could transform $A$ into a tridiagonal form is impossible. This problem is circumvented by the bi-Lanczos algorithm by evolving two initial vectors, $|q_1 \rangle$ and $\langle p_1 |=|q_1 \rangle^\dagger$ by $A$ and $A^\dagger$, instead of only $|q_1 \rangle$ as in the usual Lanczos algorithm. The initial state $|\psi(t=0)\rangle$ is chosen as $|q_1 \rangle$.   From these two initial vectors, by using the two-sided Gram-Schmidt procedure, a pair of bi-orthogonal bases are built, $\{ \langle p_j | \}$ and $\{ | q_j \rangle \}$, which span the Krylov subspaces $\mathcal{K}^j(A,| q_1 \rangle) \equiv \{ A | q_1 \rangle, A^2 | q_1 \rangle, \dots \} $ and $\mathcal{K}^j(\langle p_1 |, A) \equiv \{ \langle p_1 |A, \langle p_1 | A^2, \dots \}$ respectively.

The usual Lanczos algorithm is realized by the action of a non-hermitian matrix $A$ on a ket vector $|q_j\rangle$ from the left.  The action of the non-hermitian matrix $A$ on a bra vector $\langle p_j |$ can be realized as $A^\dag$ acting on a ket vector $|p_j\rangle$. These bases, which span the two Krylov subspaces are bi-orthogonal to each other, i.e. 
\begin{equation}
    \langle p_i | q_j \rangle \ = \delta_{ij}. \label{basis}
\end{equation}
 Since $|\psi(t=0)\rangle=|p_1\rangle = |q_1 \rangle$ is a normalised vector,   $\langle p_1 | q_1 \rangle=\langle q_1 | q_1 \rangle =1$ by definition. The action of the non-hermitian matrix $A$ or $A^\dag$ on general basis vectors, $|q_j \rangle$ or $|p_j \rangle $, makes them imaginary for $j>1$. This makes $\{ \langle p_j | \}$ and $\{ | q_j \rangle \}$ in general not orthonormal, i.e.
\begin{equation}
     \langle p_j | p_j \rangle  \neq 1 \neq \langle q_j | q_j \rangle \, \text{for} \, j > 1. \end{equation}
 Starting from the first two initial vectors $|p_1\rangle$ and $|q_1\rangle$, the two sets of vectors are generated by implementing three terms of recursive relations given by
\begin{align}
   c_{j+1}\, |q_{j+1} \rangle &= A\,  |q_j\rangle - a_j \, |q_j\rangle - b_j\,  |q_{j-1}\rangle \label{recursion1} \\ 
      b^*_{j+1} \, |p_{j+1} \rangle &=  A^{\dagger} \, |p_{j} \rangle   -  a_j^*  \, |p_{j} \rangle -  c^*_j \, |p_{j-1} \rangle  \label{recursion2}.
\end{align}    
 It is necessary to verify that each vector of Ket basis is bi-orthogonal with respect to the previously constructed vectors of the Bra basis after every iteration of the process. In this bi-orthogonal basis the matrix $A$ takes the tri-diagonal form $T_j$,
 \begin{equation}\label{eq:tj_matrix}
     T_j = \begin{pmatrix}
    a_1 & b_2 & 0        & \dots     & 0   \\
     c_2 & a_2           & b_3   &    &  \vdots  \\
     0 &  \ddots     &    \ddots      & \ddots    &     \\
\vdots &       & c_{j-1}  & a_{j-1}    & b_j   \\
     0 & \dots  &     0   &     c_j   & a_j
  \end{pmatrix}. 
\end{equation}
The recursion relations given in Eq.~\eqref{recursion1} and Eq.~\eqref{recursion2} make it obvious that this approach requires both $A$ and $A^{\dagger}$ to act on the vectors. This method is simple because it only uses two three-term recursive relations.

  Adapting the procedure given in equations \eqref{basis}-\eqref{eq:tj_matrix} to non-hermitian systems gives rise to an algorithm for the spread complexity as follows: starting from two initial vectors chosen as the initial state $|\psi(t=0)\rangle=|p_1\rangle = |q_1 \rangle$,  through iterative application of the Hamiltonian and a particular orthogonalization process, we systematically construct a bi-orthogonal set of basis vectors spanning the bi-orthogonal Krylov spaces by the following way:
\begin{enumerate}
    \item Choose two initial normalised vectors $ |p_1 \rangle$ = $|q_1\rangle =|\psi(t=0)\rangle$ such that $\langle p_1 | q_1 \rangle = 1$. Here to initiate the algorithm, we also need an initial set of values for the diagonal and off-diagonal bi-Lanczos coefficients.
    \begin{equation}
        b_1=c_1=0,~~ a_1=\langle p_1|A|q_1\rangle= \langle q_1|A|q_1\rangle. 
    \end{equation}
    \item  Starting from the initial vectors $|p_1\rangle$ and $|q_1\rangle$ which correspond to $j=1$, we implement the steps (a) to (f) given below to generate further basis vectors of the two Krylov spaces, $|p_j\rangle$ and $|q_j\rangle$, for $j>1$.  
   \begin{enumerate}
       \item   The action of $A$ on $|q_j\rangle$ and, similarly $A^{\dagger}$ on $|p_j\rangle$ gives rise to a set of vectors denoted by $|r'_j\rangle$  and $|s'_j\rangle$ respectively. 
       \begin{equation}
           |r'_j\rangle = A |q_j \rangle,  \,  \, | s'_j \rangle =  A^{\dagger} |p_j \rangle.
       \end{equation}      
   \item  The vectors $|r_j\rangle$ and $|s_j\rangle$ correspond to the right hand side of equations \eqref{recursion1} and \eqref{recursion2}. Now  in order to construct orthogonal basis vectors, we subtract contributions from two vectors of previous basis  $|q_j\rangle,\, |q_{j-1}\rangle$ and $|p_j\rangle,\, |p_{j-1}\rangle$. We obtain 
   \begin{equation}
       |r_j\rangle =  |r'_j \rangle - a_j |q_j\rangle - b_j |q_{j-1} \rangle, \, |s_j \rangle = |s'_j \rangle -  a_j^* |p_j\rangle - c^*_j |p_{j-1}\rangle, \label{eq:bilanc_ieration}
    \end{equation}
    with complex Lanczos coefficients $a_j,\, b_j,$ and $c_j$.  Their values are known up to the level $j$.
   \item In order to obtain the Lanczos coefficients of the next step, we evaluate the inner product of vectors constructed in the previous one by $\omega_j = \langle r_j | s_j \rangle$. We  define the upper and lower diagonal coefficients $b_{j+1}$ and $c_{j+1}$ for the $(j+1)$st basis vector of the $T_j$ matrix given in \eqref{eq:tj_matrix},
    \begin{equation} \label{eq:cnbn}
     c_{j+1} = \sqrt{| \omega_j |},\, \, b_{j+1} = \frac{\omega^*_j}{ c_{j+1}}.
     \end{equation}
   \item We proceed to construct the unit basis vectors $|q_{j+1}\rangle$ and $|p_{j+1}\rangle$ of the two Krylov bases,
   \begin{equation}
       |q_{j+1} \rangle = \frac{|r_j \rangle}{c_{j+1}} ,\, \,
    |p_{j+1} \rangle = \frac{|s_j \rangle}{b^*_{j+1}}.
    \end{equation}  
    \item  In any Lanczos basis construction, full orthonormality can be lost due to numerical instability caused by the finite-precision arithmetic \cite{Parlett1979,Simon1984}. This happens because the contributions of only the previous two basis vectors, instead of all previous basis vectors, are subtracted while constructing a new basis vector (see Eq.~\eqref{eq:bilanc_ieration}). The resulting errors accumulate with increasing steps causing a loss of orthonormality. This problem can be avoided by the full Gram-Schmidt orthogonalization procedure that subtracts the contributions of all previous basis vectors while constructing the new basis vector. To avoid this problem for the bi-Lanczos algorithm, we implement the full bi-orthogonalization to ensure that Eq.~\eqref{basis} is valid for all $i,j$, 
    \begin{equation}
        |q_j \rangle = |q_j \rangle - \sum_{l=1}^{j-1} \langle p_l|q_l \rangle |q _l \rangle , \,
    \, |p_j \rangle = |p_j \rangle - \sum_{l=1}^{j-1} \langle q_l|p_l \rangle |p _l \rangle. 
    \end{equation}    
\item Finally, after full bi-orthogonalization, we compute the diagonal coefficients $a_{j+1}$ of the tri-diagonal matrix for $(j+1)$st step. In usual Lanczos, this is just the expectation value of the operator $A$ with respect to the newly constructed basis vector $|q_{j+1}\rangle$. In bi-Lanczos algorithm, the modified definition of $a_{j+1}$ is given by sandwiching $A$ between the $\langle p_{j+1}|$ `bra' vector and the $|q_{j+1}\rangle$ `ket' vector,
\begin{equation}
    a_{j+1} = \langle p_{j+1} | A | q_{j+1} \rangle \label{eq:an}
\end{equation}
 and by returning to step $2$ for level $j+2$. 
\end{enumerate}
   \item If $\omega_j = 0 $ at $j= \mathcal{K}$, we end the recursion and obtain the $\mathcal{K}$-dimensional Krylov space $\mathcal{K}$ Krylov basis vectors. Here, $\omega_{\mathcal{K}}=0$ implies that no further linearly independent basis vectors are left to be formed. This happens when the Krylov basis vectors completely explore the full Hilbert space.  Then, for a $N \times N$ non-hermitian matrix A, at the $\mathcal{K}^{th}$ step, two $N \times \mathcal{K}$ matrices $ P_\mathcal{K} \equiv [p_1, p_2, \dots p_\mathcal{K} ]$ and $Q_\mathcal{K} \equiv [ q_1, q_2, \dots q_\mathcal{K}]$ are formed by bi-Lanczos method such a way that, $P^\dag_\mathcal{K} A Q_\mathcal{K} = T_\mathcal{K}$.
\end{enumerate}

This algorithm provides all the basis vectors $|p_j\rangle$, $|q_j\rangle$ of the full Krylov bases. In addition, we obtain the list of Lanczos coefficients $a_j$ Eq.~\eqref{eq:an}, $b_j$, and $c_j$ Eq.~\eqref{eq:cnbn}, which are the diagonal, upper diagonal, and lower diagonal elements of the  tri-diagonal form of $A$ mentioned in Eq.~\eqref{eq:tj_matrix} respectively.

From this algorithm it is clear that all $c_j$'s are real since $c_j = \sqrt{|\omega_{j-1}|}$, but $b_j$ can be complex Eq.~\eqref{eq:cnbn}. In general $c_n \neq b_j$ while $|c_j|= |b_j|$ \cite{Bhattacharya:2023zqt}. We may expand the wave function $|\psi(t)\rangle$ in both of the bases,
\begin{equation}
    \label{eq:components}\sum_{j=1}^\mathcal{K}\Phi^q_j(t)|q_j\rangle=|\psi(t)\rangle=\sum_{j=1}^\mathcal{K}\Phi^p_j(t)|p_j\rangle,
\end{equation}
where $\mathcal{K}$ is the dimension of the bi-orthogonal Krylov basis. 
In the bi-orthogonal basis, the corresponding amplitudes of the wave function $\ket{\psi(t)}$ have two components, $\Phi^p_j(t)$ and $\Phi^q_j(t)$ for the $|p_j\rangle$ and $|q_j\rangle$ bases respectively Eq.~\eqref{eq:components}. Using the normalisation \eqref{norm_in_kry} for each of these components, the total probability in the bi-orthogonal Krylov basis is defined by\footnote{Here we introduce the superscript (bl) to specify that these definitions are specific to the bi-Lanczos algorithm while the tildes mean that we have performed the normalisation of the coefficients to make the total probability $1$ at all times.}
\begin{equation}
    P^{(bl)}(t)=\sum_n |\tilde{\Phi}^{p*}_n(t) \tilde{\Phi}^q_n(t)|.
\end{equation}
The corresponding definitions of spread complexity and spread entropy are
\begin{align}
    C^{(bl)}_S(t)&=\sum_n n|\tilde{\Phi}^{p*}_n(t) \tilde{\Phi}^q_n(t)|,\\
    S^{(bl)}_S &=- \sum_n \left(|\tilde{\Phi}^{p*}_n\left(t\right) \tilde{\Phi}^q_n\left(t\right)|\right) \ln [|\tilde{\Phi}^{p*}_n\left(t\right) \tilde{\Phi}^q_n\left(t\right)|].
\end{align}
The bi-Lanczos algorithm and the above definitions of probability, complexity and entropy apply to any general non-hermitian operator acting on the state vectors.  However, for large Hilbert space dimensions, this method has high computational memory requirements. For this reason, we propose an alternative approach for the complex symmetric Hamiltonians below.


\subsubsection{Complex symmetric Lanczos algorithm}\label{complex-symmetric}
 
Here implement an alternative cost-efficient algorithm for complex symmetric matrices that we will study in section \ref{sec:first-passage} when considering a complex symmetric Hamiltonian $H_{\text{eff}}^{T} = H_{\text{eff}}$ but $H_{\text{eff}}^{\dag} \neq H_{\text{eff}}$. 
 
  The property of complex symmetry is purely algebraic, exerting no impact on the matrix spectrum. In general, for any given set of $n$ numbers, there exists a complex symmetric $n \times n$ matrix $A$, whose eigenvalues precisely match the prescribed numbers.  A complex symmetric matrix may not always be diagonalizable. While the complex symmetry of matrix $A$ does not put any constraints on its eigenvalues, this specific algebraic property can be used to significantly reduce the computational workload and storage demands associated with the general non-Hermitian bi-Lanczos method.

 The diagonalizability of a complex symmetric matrix, $A$ depends upon the possibility of choosing its eigenvector matrix, $Z$, in such a manner that, $Z^TAZ=\text{diag}(\lambda_1,\lambda_2,...,\lambda_n)$, where $\{\lambda_1,\lambda_2,...,\lambda_n\}$ is the set of its eigenvalues. The eigenvector matrix $Z$ additionally satisfies, $Z^TZ=I_n$ which means the $Z$ matrix is complex orthogonal. The complex orthogonality arising from the inherent complex symmetric nature of the matrix is pivotal. It allows the construction of a Krylov basis for such matrices in an efficient manner. To achieve this, we employ a modified algorithm based on the conventional Lanczos method. Even when dealing with complex vectors, we use the fact that the new Krylov basis should maintain complex orthogonality. Consequently, the Lanczos vectors constructed exhibit complex orthogonality.
 
 Utilising these properties of complex symmetry, the tri-diagonal form of a complex symmetric matrix $A$ can be obtained using the \textit{complex symmetric Lanczos} algorithm \cite{200SIAM}. The orthogonal basis $\{  | q_j \rangle \}$ that spans the Krylov space $\mathcal{K}^j (A, |q_1 \rangle) \equiv \{A | q_1 \rangle, A^2 | q_1 \rangle, \dots  \} $ is constructed starting with a normalized vector $|q _1 \rangle=|\psi(0)\rangle$. The complex orthogonality of the vectors in this situation means that the product of a vector with its transpose is $1$, i.e., $ \langle q_j |q _j \rangle = \delta_{i,j} $ where in our notation $\langle q_j | = (|q _j \rangle )^T $. The construction of the required basis $\{ |q_j \rangle \}$ involves a three-term recursion relation, 
\begin{align}
    \beta_{j+1} |q_{j+1} \rangle = A |q_{j} \rangle - \alpha_j |q_{j} \rangle - \beta_{j} |q_{j-1} \rangle . \label{recursion3}
\end{align}
 In this basis $\{ | q_j \rangle \}$, the complex symmetric matrix $A$ takes the tri-diagonal form $\Tilde{T}_j$, 
\begin{equation}
\Tilde{T}_j = \begin{pmatrix}
    \alpha_1 & \beta_2 & 0        & \dots     & 0   \\
     \beta_2 & \alpha_2           & \beta_3   &    &  \vdots  \\
     0 &  \ddots     &    \ddots      & \ddots    &     \\
\vdots &       & \beta_{j-1}  & \alpha_{j-1}    & \beta_j   \\
     0 & \dots  &     0   &     \beta_j   & \alpha_j
  \end{pmatrix}.\label{eq:ttilde_matrix}
\end{equation}\label{T_tilde}
 Using the usual Lanczos algorithm with complex orthogonality, we obtain the set of Krylov basis matrix, $Q=[\hat{q}_1~\hat{q}_2~...~\hat{q}_n]$ that satisfies $Q^TQ=I_n$. This yields the tri-diagonal form of the matrix, $A$ in this basis,  $\Tilde{T}_j=Q_j^TAQ_j$. 

Now we compare between the tri-diagonal matrices $T_j$ and $\Tilde{T}_j$ constructed using the bi-Lanczos and the complex symmetric Lanczos methods given in equations \eqref{eq:tj_matrix} and \eqref{eq:ttilde_matrix} respectively. An obvious advantage of the complex symmetric Lanczos algorithm is that we just deal with a single set of  Krylov basis vectors, $|q_j\rangle$ (Eq.~\eqref{recursion3}). In the complex symmetric Lanczos method, upper and lower diagonal elements of the tri-diagonal matrix $\Tilde{T}_j$ are the same. This makes the $\Tilde{T}_j$ symmetric. On the other hand, $T_j$ obtained by the bi-Lanczos algorithm is not symmetric (Eq.~\eqref{eq:tj_matrix}). The diagonal elements of $T_j$ and $\Tilde{T}_j$ are the same i.e., $a_j = \alpha_j$. The $j^{th}$ upper as well as the lower diagonal elements  of $T_j$ and $\Tilde{T}_j$ are not the same i.e., $b_j \neq \beta_j$ also $c_j \neq \beta_j$. However, absolute values are the same i.e., $|\beta_j| = |b_j| = |c_j|$. Hence $|T_j| = |\Tilde{T}_j |$.

This algorithm for complex symmetric matrices will be of use below in section \ref{sec:setup_and_results} where we study Hamiltonians of a complex symmetric form, in particular for the QFPP.

 
\section{Complexity and entropy under projective measurements}\label{sec:setup_and_results}

To study the different features of non-unitary evolution, we focus on a particular example known as the ``quantum first passage problem". The methods developed in the previous section are generic and applicable to any non-unitary evolution, whereas the results of this section are problem-specific. The problem-specific results indicate how sensitive the Krylov basis measures are to specific properties of the QFPP. We first review the first passage problem in section \ref{sec:first-passage}. Then, we discuss the numerical findings of this non-unitary evolution in section \ref{sec:results}.

\subsection{First passage problem}\label{sec:first-passage}
The quantum first passage problem (QFPP) in quantum mechanics consists of determining the probability that a particle starting from a specific initial position reaches a particular final position for the first time during evolution within a given time \cite{muga2000}. The QFPP is relevant in different contexts, even beyond physics, including the study of chemical reactions, electron transport in materials, and the behaviour of particles in biological systems. 

In \cite{Dhar_2015, Dhar2013QuantumTO}, it was shown that the time evolution of a quantum mechanical system under periodic projective measurements may be viewed as a non-hermitian system with a non-unitary time evolution. We use the setup presented in \cite{Dhar_2015, Dhar2013QuantumTO} as our prime example here.  This setup involves a quantum particle moving on a lattice governed by a tight-binding Hamiltonian. The position of the particle is periodically measured to determine whether the particle is present at specific predefined sites. We begin by considering the unitary time evolution of the wave function between the points at which detection occurs. Projective operations represent the detection procedures, giving rise to non-unitarity. 

We consider a system undergoing unitary time evolution subjected to repeated projective measurements, effectively giving rise to the non-unitary evolution of the system. In the most general case, it is important to note that this system has a complex eigenspectrum without a complex conjugate pair of eigenvalues. In general, we start with a Hamiltonian of the form,
\begin{equation}
    H=\underbrace{\sum_{l,m}H_{l,m}|l\rangle\langle m|}_{H_S}+\underbrace{\sum_{\alpha,\beta}H_{\alpha,\beta}|\alpha\rangle\langle \beta|}_{H_M}+\underbrace{\sum_{\alpha,l}\left(V_{\alpha,l}|\alpha\rangle\langle l|+V_{l,\alpha}|l\rangle\langle \alpha|\right)}_{V},
    \label{Eq:genhamiltonian}
\end{equation}
where each state $|l \rangle$ represents a specific location on the lattice. The complete set of sites is divided into two categories: those associated with the system, designated by Roman indices $l$ and $m$,  and those comprising the domain of sites where measurements are conducted, denoted by Greek indices $\alpha$ and $\beta$. In accordance with the notation put forth in previous works \cite{Dhar_2015}, the terms in Eq.~\eqref{Eq:genhamiltonian} are consecutively referred to as $H_S$, $H_M$, and $V$. The unitary time evolution is 
\begin{equation} 
|\psi(t)\rangle = U_t |\psi(0)\rangle ,\, \, \text{with}\, \, U_t = e^{-iHt}. 
\end{equation} 
In order to take into account the projective measurements, we define the projection operator defined at detection sites as $A= \sum_{\alpha\in D}|\alpha\rangle\langle \alpha|$. This operator corresponds to a measurement for detecting the particle within domain $D$, which contains a specific number of sites. The complementary operator $B=1-A$ corresponds to the projection onto the sites of the system subspace. 
 The probability $p$ of detecting the particle on measuring the state $|\psi\rangle$ can be found by calculating the expectation value of $A$ concerning that state. Subsequently, the probability of that state not being detected or the so-called {\it survival probability} defined in \cite{Dhar_2015} is given by $P = \langle\psi|B|\psi\rangle = \left(1-p\right)$. In the case of a positive detection outcome, the state right after the measurement is $A|\psi\rangle$.
    On the other hand, if the state is not detected right after the measurement, the state must have collapsed to $B|\psi\rangle$. We repeat this procedure indefinitely until the state is finally detected.

 In this setup, the effective time evolution between two consecutive measurements is of the form 
\begin{equation}
\Tilde{U}_{\tau}=B\cdot e^{-iH\tau}\cdot B = B \cdot U_{\tau} \cdot B   .
\end{equation}
Here, $\tau$ is the time interval between two consecutive measurements. The total time evolution of the system is given by
\begin{align}
   \mathcal{U}_{M+1}=\left( \Tilde{U}_{\tau}\right)^M=B\cdot U_{\tau}\cdot B\cdot U_{\tau}\cdot \cdot \cdot  \underbrace{B\cdot U_{\tau}}_{M~\text{times}}\cdot B,
\end{align}
where in the last step, we have used $B^2=B$. The operator $\mathcal{U}_{M+1}$ acts as the time evolution operator for $M+1$ number of projective measurements. $B$ being not unitary leads to an effective non-hermitian Hamiltonian through second-order perturbative calculations for small $\tau$. This effective Hamiltonian is given by,
\begin{align}
    \Tilde{U}_{\tau}&=e^{-iH_{\text{eff}}\tau}+\mathcal{O}(\tau^3)\rightarrow\mathcal{U}_{M+1}\approx e^{-i H_{\text{eff}}t}~~\text{with}~~t=M\tau,~~\\\text{where}~~H_{\text{eff}}&=H_S-\frac{i\tau}{2}\sum_{l,m}\sum_\alpha V_{l,\alpha}V_{\alpha,m}|l\rangle\langle m|.\label{eqtau}
\end{align}
We notice that $H_{\text{eff}}$ is a linear combination of a hermitian part $H_S$ and a non-hermitian part. This effective Hamiltonian acts on all the sites except for the sites where the detectors are present.\footnote{From now onwards, we refer to the system as the chain without the detector sites.} The second term in Eq.~\eqref{eqtau} determines the complex part of the eigenspectrum. The first hermitian part always contributes to the real part of the eigenvalues. It is important to note that the limit $\tau\rightarrow 0$ does not coincide with setting $\tau =0$ in the effective non-hermitian Hamiltonian from the beginning.  This is because the effective Hamiltonian is derived from a perturbative expansion. In the mathematical expression, $\tau = 0$ gives back the hermitian Hamiltonian, whereas the physical limit $\tau\rightarrow 0$ represents almost continuous measurements. Additionally, interpreting $\tau\rightarrow\infty$ as the unitary limit, where the time gap between measurements becomes very large, is not accurate. In this scenario, the perturbative treatment breaks down.

We consider two one-dimensional models, i) one with open boundary conditions and ii) one with periodic boundary conditions. We start with the initial tight-binding Hamiltonian of $N$ sites,
\begin{equation}
    H_{\text{TB}}=-\sum_{l=1}^{N-1}\left(|l+1\rangle\langle l|+|l\rangle\langle l+1|\right)
    \label{eq:tbham_herm}
\end{equation}
and place the detector at the end, on the $N$th point. So, for this particular model, the operator $A$ takes the form, $A=|N\rangle\langle N|$ and $B$ operator is given as $B=\sum_{l=1}^{N-1}|l\rangle\langle l|$. Using these, we can get the $H_{\text{eff}}$ for this model up to first order in $\tau$ as,
\begin{equation}
    H_{\text{eff}}=-\sum_{l=1}^{N-2}\left(|l+1\rangle\langle l|+|l\rangle\langle l+1|\right)-\frac{i\tau}{2}|N-1\rangle\langle N-1|
     \label{eq:tbham_non-herm_open}.
\end{equation}
The second term on the right-hand side represents the $V_{\text{eff}}$ that is non-hermitian. In this case, the endpoints of the chain are open. But we can also use periodic boundary conditions as $|N+1\rangle=|1\rangle$. In this case, the operator $A$ and $B$ are still the same, but the effective Hamiltonian changes as the $V_{\text{eff}}$ is different. In this case, we find
\begin{equation}
    H_{\text{eff}}=-\sum_{l=1}^{N-2}\left(|l+1\rangle\langle l|+|l\rangle\langle l+1|\right)-\frac{i\tau}{2}\left(|N-1\rangle\langle N-1|+|1\rangle\langle 1|+|1\rangle\langle N-1|+|N-1\rangle\langle 1|\right).
         \label{eq:tbham_non-herm_periodic}
\end{equation}
The periodic boundary conditions require some attention. This Hamiltonian has $(N-2)/2$ eigenvalues with two-fold degeneracy. In this case, following the form of the energy eigenstates discussed in \cite{Dhar_2015}, half of them remain unaffected by these measurement processes. In contrast to open boundary conditions, the survival probability does not decay to zero at late times unless the initial state is localised at the $(N/2)^{th}$ site for an even number of $N$ \cite{Dhar_2015}. However, the survival probability decays to zero at late times only if the initial state is localised at the $(N/2)^{th}$ site.

\subsection{Results for Krylov state complexity and entropy}\label{sec:results}
The total Krylov probability, as defined in section \ref{nhnewnorm}, remains constant even for non-unitary evolution. However, in Appendix \ref{app:K_probability}, we present an additional definition of total probability more commonly used in the literature, which is time-dependent for non-unitary evolution.  Here, we follow the normalized spread complexity and spread entropy as given in Eq.~\eqref{entcomp}.

We now proceed to discuss the numerical results for the spread complexity and entropy exhibited by the first passage problem. The presentation of our findings is divided into three distinct parts. Initially, we examine the spread complexity of the non-hermitian tight-binding chain, considering both i) open (section \ref{obc}) and ii) periodic boundary conditions (section \ref{pbc}). This analysis encompasses different values for the non-Hermiticity perturbation parameter $\tau$, as well as different positions and spreads of the initial state. Subsequently, in iii) we shift our focus to the investigation of Krylov spread complexity within a specific quench scenario (section \ref{quenched}), where we initialise the system with the eigenstates of the hermitian tight-binding Hamiltonian and evolve it using the effective non-hermitian Hamiltonian. In addition to the Krylov complexity, we also explore the behaviour of the Krylov entropy and total probability across all aforementioned systems, leading us to identify the characteristics attributed to the non-hermitian nature of the system. 

\subsubsection{Open boundary conditions}\label{obc}
We start with a tight-binding chain with open boundary conditions, under the repeated projective measurements whose effective time evolution is governed by the effective non-hermitian  Hamiltonian given in Eq.~\eqref{eq:tbham_non-herm_open}. In the following subsections, we first discuss the time evolution of the actual wave function and how it reaches a steady state. Then we discuss the time evolution of the spread complexity and spread entropy.

\subsubsection*{Steady state}\label{sec:steadystate_obc}
\begin{figure}[hbtp]
	\centering
	\includegraphics[width=0.7\textwidth]{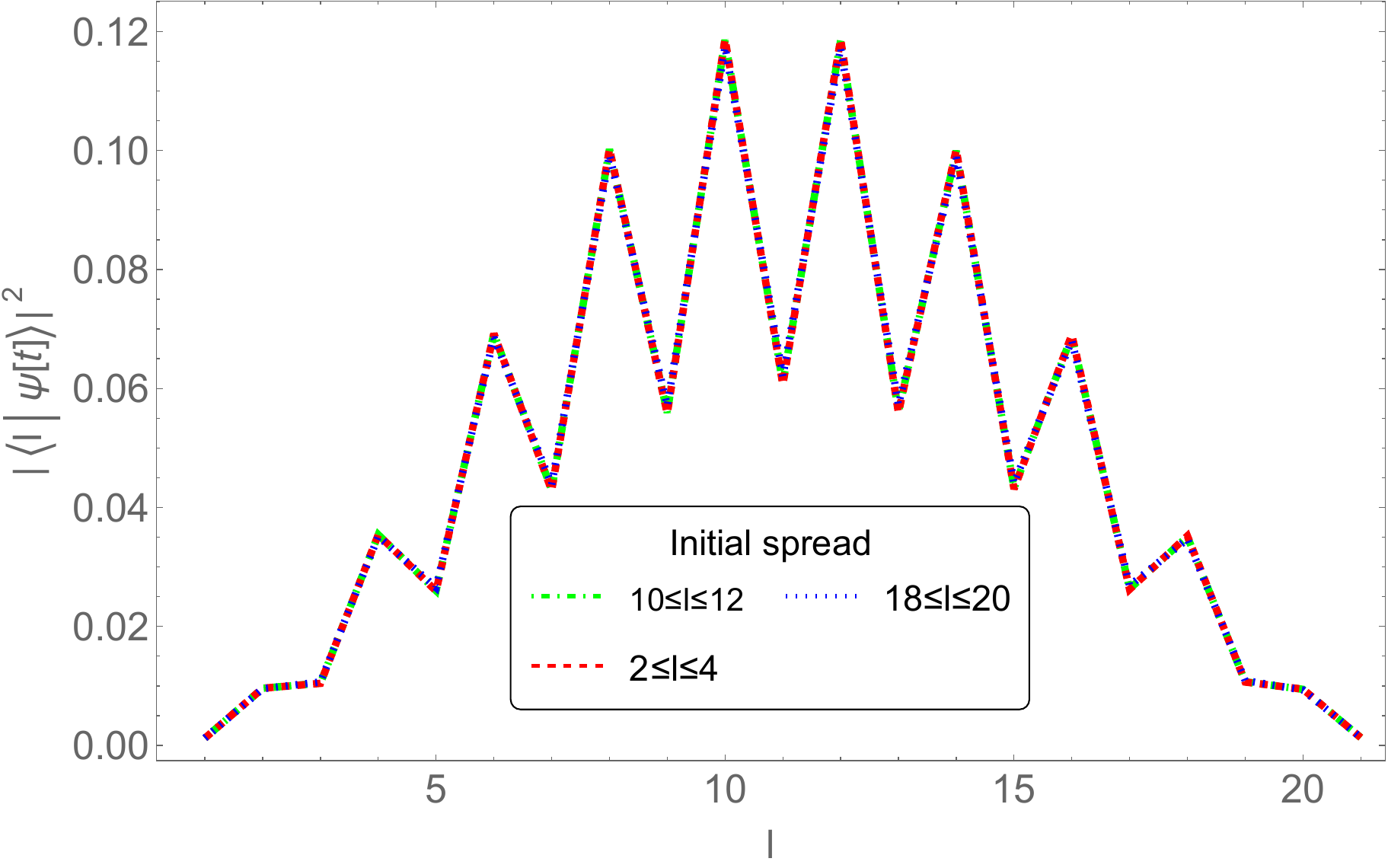}
	\caption{Spread of the steady state at late times $t=30000$ for different spread of initial state, for open boundary conditions with total number of sites $N =22$ and non-hermiticity parameter $\tau = 0.1$. The spread of the steady state is the same for different initial states. }
	\label{fig:steady_state_open}
\end{figure} 
We consider the  QFPP with open boundary conditions. Steady states are those states that remain invariant up to small fluctuations under the action of the Hamiltonian. These states for non-equilibrium systems are similar in nature to the thermal states of unitary evolution. For the QFPP, as an initial state evolves following Eq.~\eqref{nhdensity}, it explores the full Krylov basis and eventually reaches a steady state. This results in the saturation of the spread complexity and entropy. For open boundary condition in the QFPP, we find that the average position of the steady state is located  at the $l$-th site of the tight-binding chain of length $N=2l$, as we see in Figure \ref{fig:steady_state_open}. The steady state profile does not depend on the position of the initial state or the time $\tau$ between two measurements. 
For the dependence of the spread complexity on  initial state chosen for constructing the Krylov basis, given in  Eq.~\eqref{eq:initialstateKry}, we find that for initial states which already have a similar spread as the steady state reached later, the spread complexity has a very low saturation value.  Under repeated action of the Hamiltonian, the Krylov space wave function explores the neighbouring sites around the initial state. This results in a strong dependence of the complexity dynamics on the choice of the initial state. 

\subsubsection*{Spread complexity}
\begin{figure}[hbtp]
     \centering
     \begin{subfigure}[b]{0.48\textwidth}
         \centering
         \includegraphics[width=\textwidth]{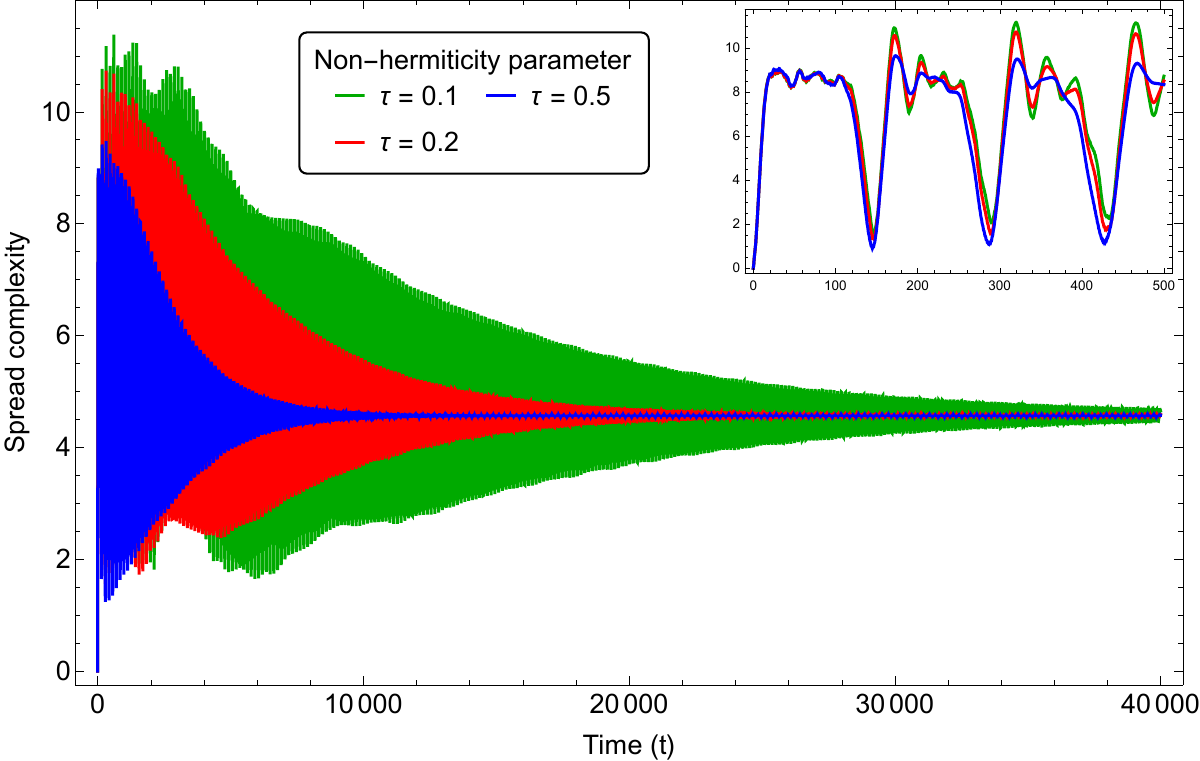}
         \caption{Time variation of spread complexity for different values of $\tau$.}
         \label{fig:kry_com_diff_tau}
     \end{subfigure}
     \hfill
     \begin{subfigure}[b]{0.48\textwidth}
         \centering
         \includegraphics[width=\textwidth]{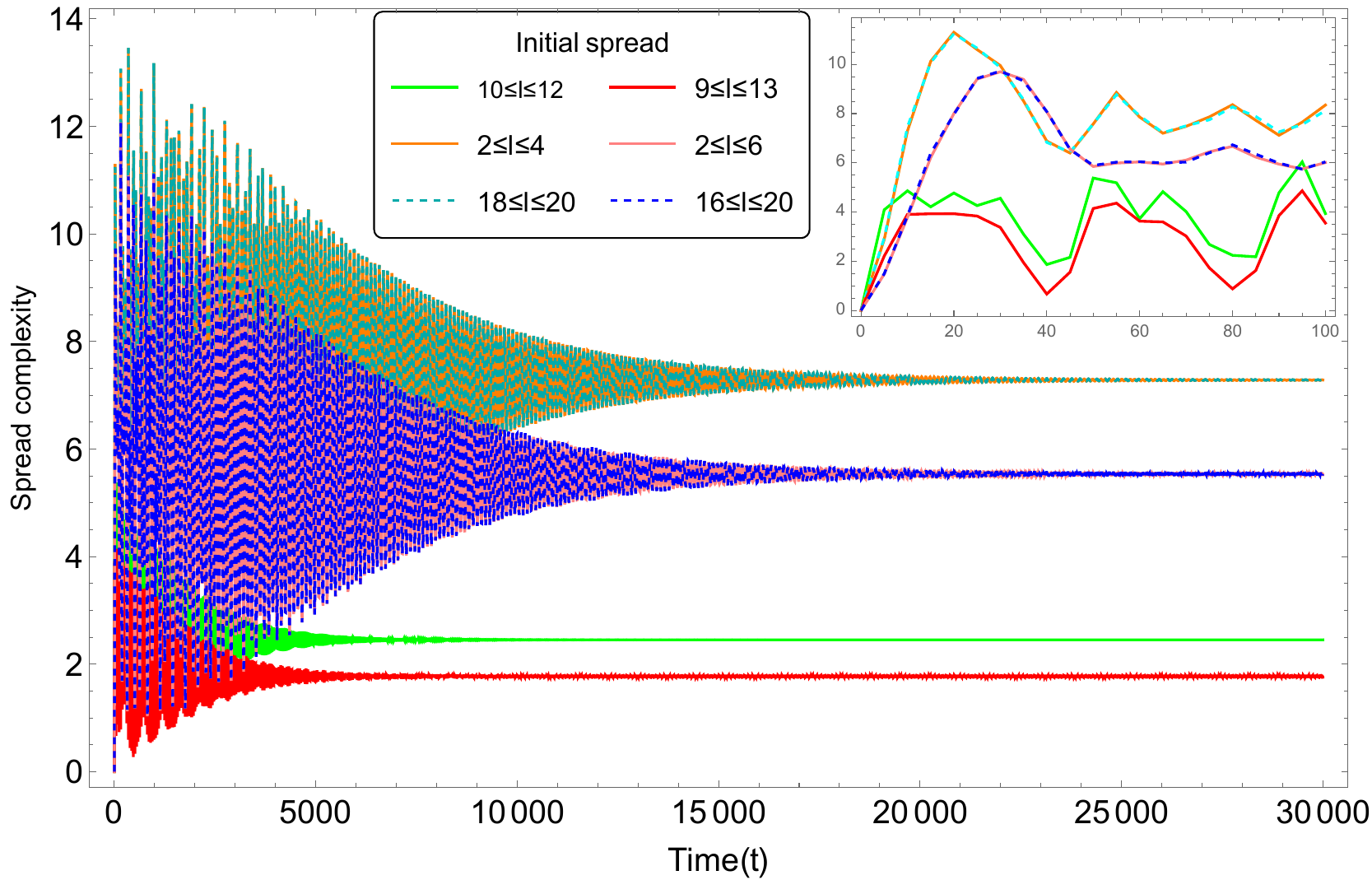}
         \caption{Time variation of spread complexity for different spreads of initial state.}
         \label{fig:kry_com_open}
     \end{subfigure}
        \caption{ (a) Time dependence of spread complexity for different non-hermiticity parameters $\tau=0.1$ (green), $0.2$ (red), and $0.5$ (blue) with fixed total sites $N=42$ and initial spread of $18\leq l \leq 22$. (b) Time dependence of spread complexity for different spreads of the initial state $9\leq l \leq 13$ (red), $10\leq l \leq 12$ (green), $2\leq l \leq 6$ (pink), $2\leq l \leq 4$ (orange), $16\leq l \leq 20$ (dashed blue), and $18\leq l \leq 20$ (dashed cyan) with fixed values of $N=22$ and $\tau=0.1$. The insets display the very early-time behaviour of the spread complexity. The decay region is prolonged for smaller $\tau$ and larger initial spread.}
        \label{fig:complexity_opn2}
\end{figure}

We plot the temporal behaviour of the Krylov spread complexity given by Eq.~\eqref{entcomp} for the tight-binding chain in Figure \ref{fig:complexity_opn2},  for different values of the non-hermiticity parameter $\tau$ in Figure \ref{fig:kry_com_diff_tau} and for different spreads of the initial state in Figure \ref{fig:kry_com_open}.  

The Krylov spread complexity initially experiences growth followed by prolonged oscillations and eventually saturates to a constant value. The amplitude of these oscillations diminishes over time, and the complexity stabilizes at a saturation point much lower than the initial peak value. The oscillatory behaviour has  different origins. Firstly, it reflects the integrability of the underlying system. Moreover, it is caused by the repeated projective measurements as well as the finite length of the chain. As the inset of Figure \ref{fig:kry_com_diff_tau} shows, changes in the measurement frequency result in a change in the oscillation period. The saturation of complexity at late times indicates a steady state as explained the paragraph above the steady states. 

For determining the dependence of spread complexity on the non-hermiticity parameter, we plot the spread complexity for the initial state spreading over sites $18\leq l \leq 22$ for different non-hermiticity parameters $\tau=0.1$ (green), $0.2$ (red), and $0.5$ (blue) in Figure \ref{fig:kry_com_diff_tau}. Note that $\tau$ corresponds to the time elapsing between two measurements. We notice that a more rapid decay of the oscillation amplitude occurs for higher values of $\tau$, resulting in saturation on a shorter time scale. This reveals that the longer the system evolves between two consecutive measurements, the quicker it reaches the steady state corresponding to the saturation. Larger time intervals between two consecutive measurements allow for the effect of the interaction to spread further into the entire system.

Moreover, it is essential to note that the saturation value of spread complexity in Figure \ref{fig:kry_com_diff_tau} remains unaffected by changes in $\tau$. It is in agreement with previous results for the open system non-unitary operator complexity  reported in \cite{Bhattacharya:2023zqt}, where changing the non-hermitian couplings in the Lindbladian does not change the saturation value of the complexity. This suggests that while a larger value of $\tau$ makes the system reach a steady state sooner, the steady state complexity for a given initial state is universal and independent of $\tau$. 


We find that the value and the rate of the saturation rely on the initial state's position and spread in the chain of the tight-binding model (see Figure \ref{fig:kry_com_open}). 
In particular, a narrower spread of the initial state leads to a higher saturation value for the spread complexity. It also reaches saturation sooner as compared to states with broader initial spreads. As is evident from Figure \ref{fig:kry_com_open}, the spread complexity corresponding to the state initially spreading over $9 \leq l \leq 13 $ (red line) takes longer to  saturate and saturates at a lower value as compared to the state initially spreading over $10 \leq l \leq 12 $ (green line). This observation indicates that for a larger spread of the initial state, the system needs more time to reach the steady state.  This steady state (Figure \ref{fig:steady_state_open}) is spread around the central site and, therefore, more complex compared to the initial states that spread away from the central site. 

If we choose the initial state spread around the centre, it can reach the steady state in less time with less saturation value of complexity. For example, in Figure \ref{fig:kry_com_open}, in comparison to the spread complexity for the initial spread over $18 \leq l \leq 20 $ (cyan dashed line), the spread complexity for the initial spread over $10 \leq l \leq 12 $ (green line) saturates quicker in time and at a lower value. This indicates that the farther the initial state is from the central site of the chain, the more complex the universal steady state is compared to the initial state.
 
The spread complexity further demonstrates a symmetric nature with spread (for example, $2\leq l \leq 4$ and $18\leq l \leq 20$) around the central site ($11$-th site) for the tight-biding chain of length $N=22$. If we begin the evolution with two initial states at the same distance from the centre of the chain and the same initial width, their spread complexities are identical.  In Figure \ref{fig:kry_com_open}, the overlap of the spread complexity for the initial spread $2 \leq l \leq 4$ ( orange line) and $18 \leq l \leq 20$ ( cyan dashed line) (also for the initial spread $2 \leq l \leq 6 $ ( pink line) and $16 \leq l \leq 20$ ( blue dashed line) clearly indicates the above statement. This behaviour can be traced back to the symmetrical nature of the survival probability $S(t)=\langle \psi(0)|\psi(t)\rangle$ in this system \cite{Dhar_2015} (see Appendix \ref{app:K_probability} for analytical expressions). The decay of the time-dependent Krylov probability described in Appendix \ref{app:K_probability} also shows an identical symmetric nature with respect to the central site for the position of the initial state. 
\subsubsection*{Spread entropy}
\begin{figure}[hbtp]
	\centering
	\begin{subfigure}[b]{0.48\textwidth}
		\centering
		\includegraphics[width=\textwidth]{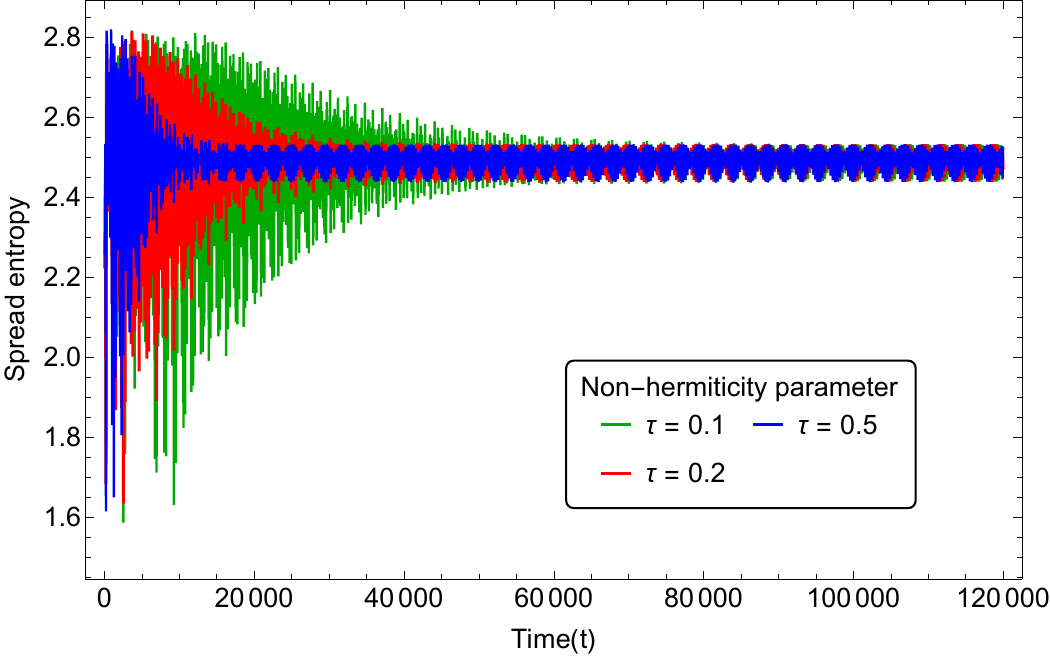}
		\caption{Time variation of spread entropy for different $\tau$.}
		\label{fig:kry_ent_diff_tau}
	\end{subfigure}
	\hfill
	\begin{subfigure}[b]{0.48\textwidth}
		\centering
		\includegraphics[width=\textwidth]{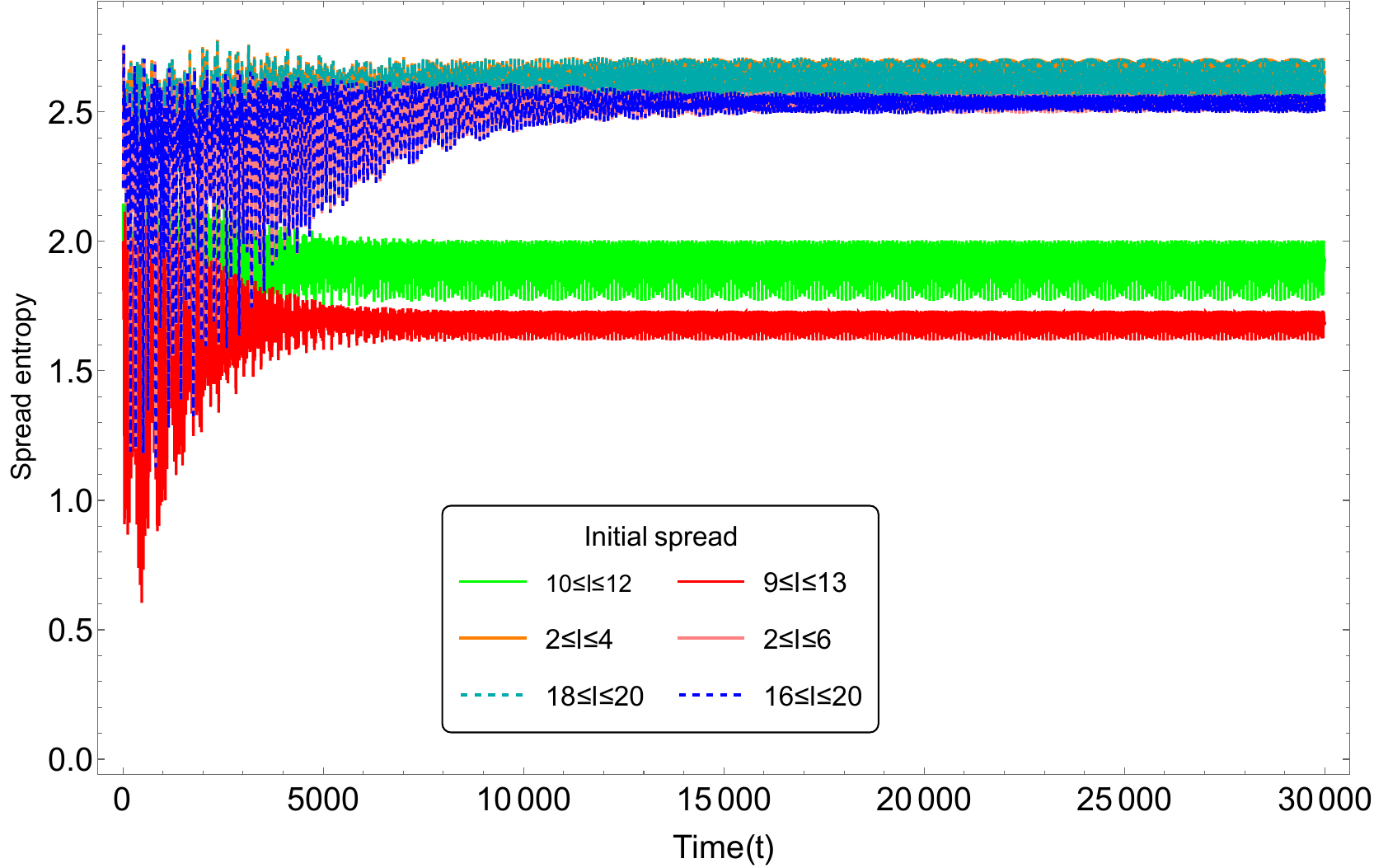}
		\caption{Time variation of spread entropy for different spreads of initial state.}
		\label{fig:kry_ent_open}
	\end{subfigure}
	\caption{ (a) Time dependence of spread entropy for different non-hermiticity parameters $\tau=0.1$ (green), $0.2$ (red), and $0.5$ (blue) with fixed total sites $N=42$ and initial spread of $18\leq l \leq 22$. (b) Time dependence of spread entropy for different spreads of the initial state  $9\leq l \leq 13$ (red), $10\leq l \leq 12$ (green), $2\leq l \leq 6$ (pink), $2\leq l \leq 4$ (orange), $16\leq l \leq 20$ (dashed blue), and $18\leq l \leq 20$ (dashed cyan) with fixed value of $N=22$ and $\tau=0.1$. We observe that spread entropy behaves in a qualitatively similar way to spread complexity (see Figure \ref{fig:complexity_opn2}).}
	\label{fig:entropy_opn2}
\end{figure}
Next, we turn to the time evolution of spread entropy, defined in  Eq.~\eqref{entcomp}, for the chain with open boundary conditions. The time dependence of the spread entropy is given in Figure \ref{fig:entropy_opn2}. We notice that spread entropy shows rapid initial growth followed by an oscillation with decreasing amplitude and, eventually, saturation.  Figure \ref{fig:kry_ent_diff_tau}, shows temporal behaviour of spread entropy for different non-hermiticity parameters $\tau=0.1$ (green), $0.2$ (red), and $0.5$ (blue). The dependence of spread entropy on the non-hermiticity parameter is qualitatively equivalent to that reported previously for the spread complexity. Higher values of $\tau$ result in a more rapid saturation because of the rapid decline of the oscillation amplitude of the spread entropy. Also, the saturation value of the spread entropy is the same for different $\tau$. This observation strengthens our previous conclusion about the universality of the steady state for a given initial state.

Next, we summarise our findings about the spread entropy for different choices of initial states. The value and rate of saturation of the spread entropy, similar to spread complexity, depend on the position and spread of the initial state. Spread entropy initially experiences growth, then oscillates, and eventually saturates at a constant value. We plot the spread entropy with time for different spread and positions of the initial state in Figure \ref{fig:kry_ent_open}. Different colour plots indicate the same configuration of the initial state as Figure \ref{fig:kry_com_open}. The spread entropy  in Figure \ref{fig:kry_ent_open} corresponding to the states initially spread around the central site ($9\leq l \leq 13$ and $10\leq l \leq 12$ for $N=22$ sites) of the chain saturate sooner to a lower value than other choices ($2\leq l \leq 4$ and $18\leq l\leq 20$). For unitary evolution, the similar behaviour of spread entropy and complexity motivated \cite{Balasubramanian:2022tpr} to conjecture that the spread complexity can be approximated by the exponential of spread entropy. Our results imply that this conjecture also applies to non-hermitian Hamiltonian evolution.

\subsubsection{Periodic boundary condition}\label{pbc}
Here, we look at the QFPP with periodic boundary conditions characterized by the effective non-hermitian Hamiltonian given in Eq.~\eqref{eq:tbham_non-herm_periodic}. In this case, we first discuss the time evolution of the wave function and characterise the profile of the steady state in the position basis. Then, we discuss the time dependence of spread complexity and spread entropy using the non-hermitian normalisation in Eq.~\eqref{newnorm} and the definitions given in Eq.~\eqref{entcomp}. We report the results for varying initial states in this section as it has important physical consequences for periodic boundary conditions, which are significantly different from the open boundary conditions.\footnote{The behaviour for varying $\tau$ is identical to the behaviour reported for open boundary conditions.} In the spread complexity and entropy profiles, the decay region is less profound if the initial states are spread near the detector. In this case, we also see more oscillations around the saturation value.

\subsubsection*{Steady states}
\begin{figure}[hbtp]
	\centering
	\includegraphics[width=0.8\textwidth]{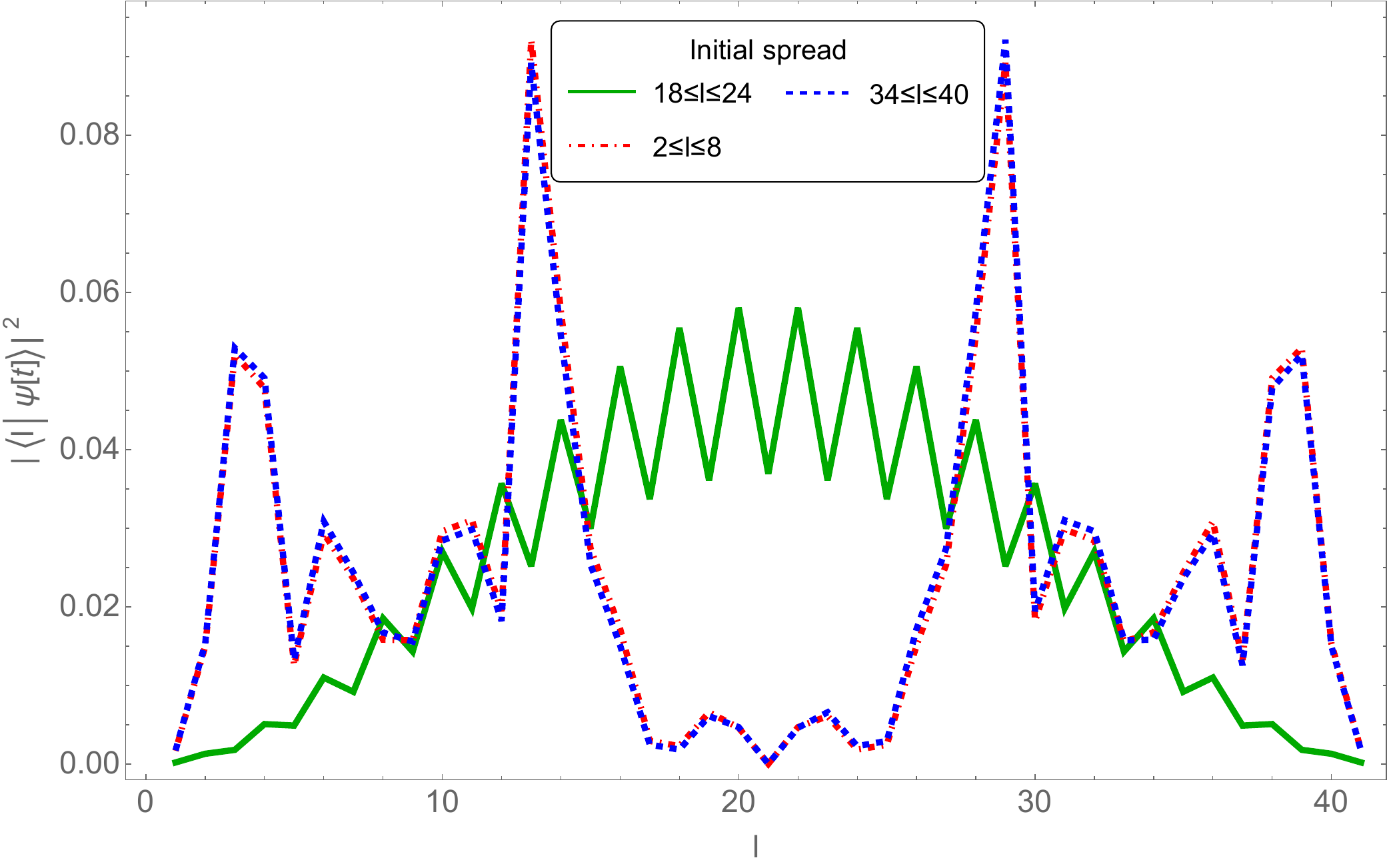}
	\caption{ Spread of the steady state at late times ($t= 50000$) for different spread of initial state for periodic boundary conditions with $N =42$ total sites   and non-hermiticity parameter $\tau = 0.1$. For periodic boundary conditions, the spread of the steady state depends on the initial state, unlike the open boundary condition case shown in Figure \ref{fig:steady_state_open}.}
	\label{fig:steady_state_periodic}
\end{figure} 
 Here we discuss the steady states for QFPP with periodic boundary conditions, evolving under the effective non-hermitian Hamiltonian given by Eq.~\eqref{eq:tbham_non-herm_periodic}. Unlike the open boundary conditions discussed in section \ref{obc}, the steady states are different for different choices of initial states. The spread of the steady state for different initial states is plotted in Figure \ref{fig:steady_state_periodic}. A localised steady state  is found only when the position of the initial state is around the central site ($18\leq l \leq 24$) of the chain of length $N=42$. 

 We find that the steady state is located at the central site of the chain only when the position of the initial state is in the middle ($l$-th) site of the lattice chain of length $2l$. However, when the initial state is spread close to the detector ($2\leq l \leq 7$ and $35\leq l \leq 40$ for $N=42$), then non-unitary time evolution does not lead to any localised steady state. The wave function is distributed with varying probability at different sites of the chain. At late times, the wave function for the initial state spread on one side oscillates around this distribution.

Next we report the behaviour of the spread complexity and entropy, which are consistent with these findings at late times.

\subsubsection*{Spread complexity}
\begin{figure}[H]
     \centering
     \begin{subfigure}[b]{0.48\textwidth}
         \centering
         \includegraphics[width=\textwidth]{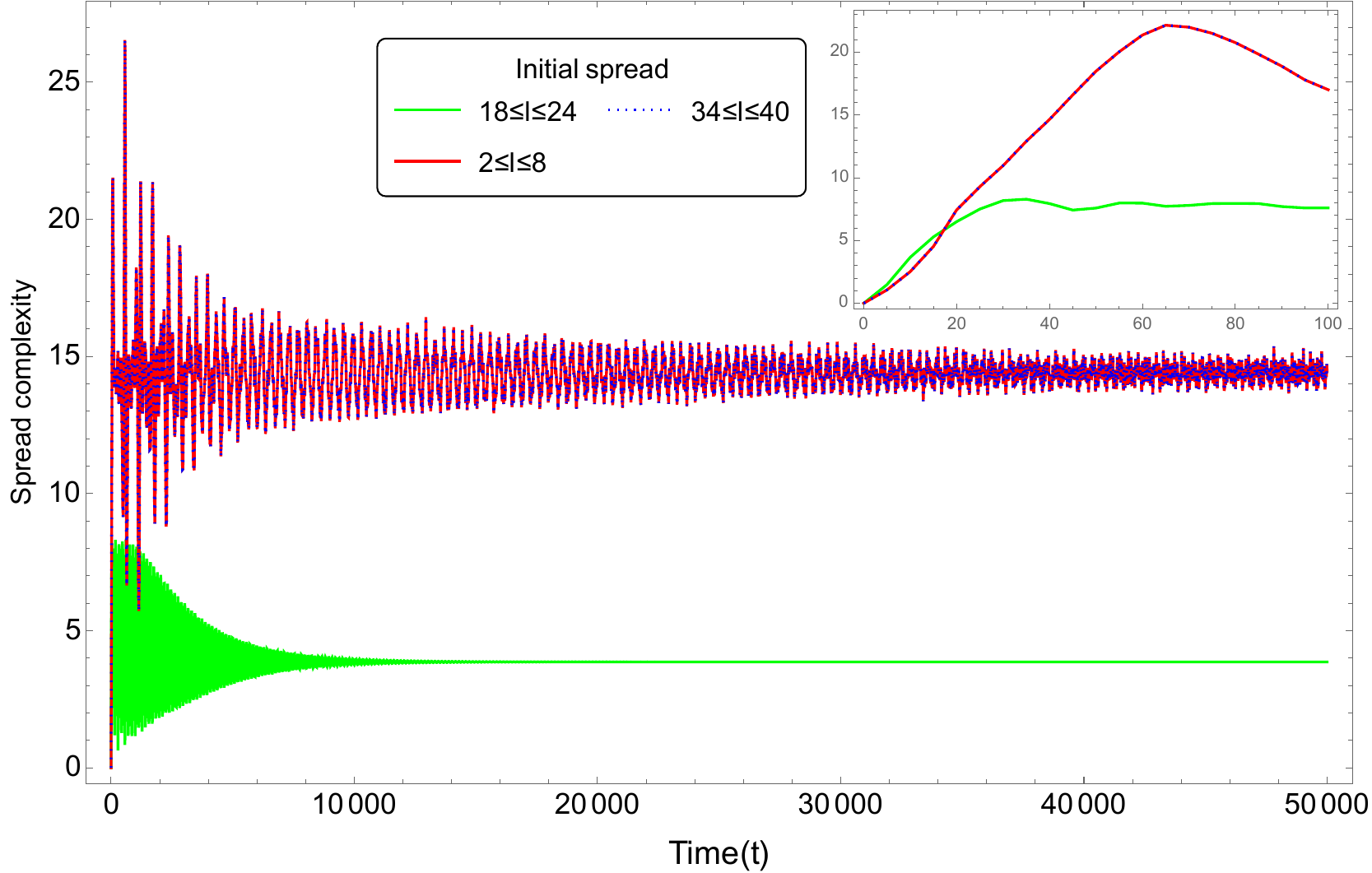}
         \caption{Time variation of spread complexity for different spreads of initial state.}
         \label{fig:kry_com_diff_spd_per}
     \end{subfigure}
     \hfill
     \begin{subfigure}[b]{0.48\textwidth}
         \centering
         \includegraphics[width=\textwidth]{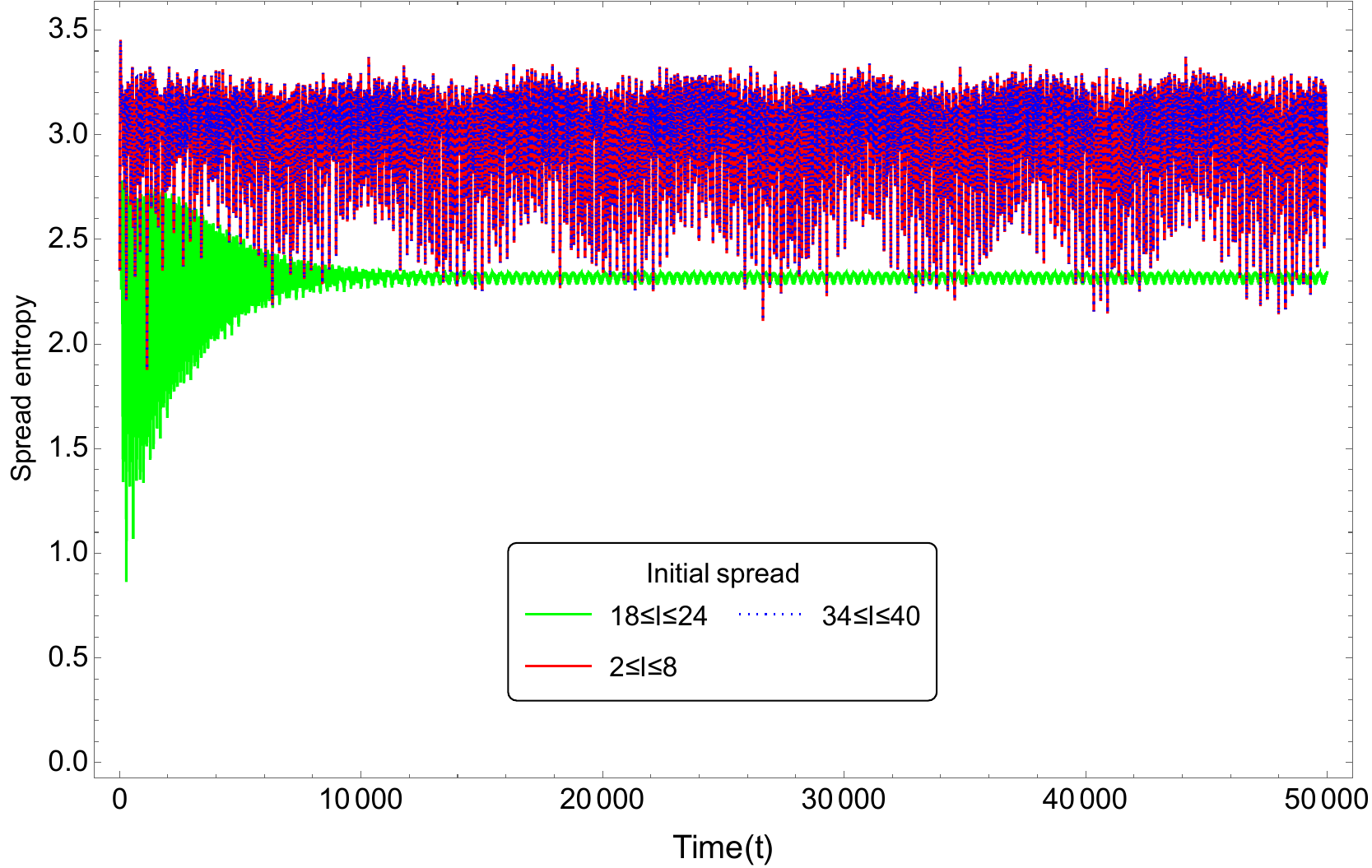}
         \caption{Time variation of spread entropy for different spreads of initial state.}
         \label{fig:kry_ent_diff_spd_per}
     \end{subfigure}
        \caption{ (a) Time dependence of spread complexity and (b) Time dependence of spread entropy for different spreads of the initial state $18\leq l \leq 24$ (green), $2\leq l \leq 8$ (red),  $34\leq l \leq 40$ (dotted blue) with fixed value of $N=42$ and $\tau=0.1$ with periodic boundary condition. For initial state spread around the centre of the chain (green), the oscillations die out quickly. For off-centre initial spread (red, dotted blue), we observe an oscillatory saturation phase.}
        \label{fig:comp_entro_periodic}
\end{figure}

 We plot the time dependence of spread complexity Eq.~\eqref{entcomp} in Figure \ref{fig:kry_com_diff_spd_per} for different spreading of initial state, keeping $N = 42$ and $\tau = 0.1$ fixed. When the initial spread includes the central site ($l$-th site for $N=2l$), the complexity shows similar growth, decay, and saturation behaviour found for the open boundary conditions. For example, see the spread complexity of the initial state with an initial spread of $18 \leq l \leq 24$ (green plot in Figure \ref{fig:kry_com_diff_spd_per}). The saturation value for this case is highly suppressed as compared to the other plots in the figure. This suppression of complexity arises from the fact that, for an initial state distributed around the centre of the chain, the dimension of the Krylov space reduces to half of the total dimension. This follows since the steady state is localised at the centre of the tight-binding chain when the initial state is spread around this centre (Figure \ref{fig:steady_state_periodic}). Therefore, the steady state is reached  sooner, and the saturation value of complexity is less. However, if the initial state is spread away from the centre of the chain, the time-evolved state never reaches a localised steady state, as shown in Figure \ref{fig:steady_state_periodic}. It oscillates near a state which has support on all sites. This is reflected by the oscillations shown in the red and blue plots for complexity and entropy in Figures \ref{fig:kry_com_diff_spd_per} and \ref{fig:kry_ent_diff_spd_per}, respectively.
  When the initial spread does not include the central site, the complexity behaves as in an effectively unitary system at late times, as indicated by a short decay period and prolonged oscillatory saturation region. As we notice, in Figure \ref{fig:kry_com_diff_spd_per}, the spread complexity associated with the initial states of spreading $ 2 \leq l \leq 8$ (red line) and $34 \leq l \leq 40$ (dotted blue line) do not have the prolonged decay region. They oscillate with a small amplitude after initial growth. However, the amplitude of the oscillation depends on the size of the spread of the initial state. The oscillation amplitude is larger for a larger spread.  

The corresponding time-dependent Krylov probability, defined under Eq.~\ref{k_com_unitary}, also exhibits signs of effective unitarity. This is discussed in detail in Appendix \ref{app:K_probability} and shown in Figure \ref{fig:Toal_prob_diff_spd_periodic}.  The total time-dependent probability stops decaying at a relatively early time and saturates at a constant value.  For the off-centre spread where the time-dependent probability does not decay completely to zero, the spread complexity plots show a higher saturation value (resulting from a less profound decay) than the cases with initial spread around the centre of the chain.

 \subsubsection*{Spread entropy}
 In Figure \ref{fig:kry_ent_diff_spd_per}, we show  the spread entropy for different initial states. The spread entropy shown here is similar to the corresponding spread complexity shown in Figure \ref{fig:kry_com_diff_spd_per}. 
For the states with the initial spread around the centre of the chain,  the spread entropy again shows large oscillations, which decrease with time and saturate to a fixed value. For the states with an initial spread near the detector, spread entropy grows and oscillates around a saturation value without showing any decay. The red and blue plots of spread entropy in Figures \ref{fig:kry_ent_diff_spd_per} and \ref{fig:kry_com_diff_spd_per} also clearly show the symmetric nature around the central site, as reported previously for steady state and spread complexity. In the spread entropy,  we also find the effective unitary behaviour for initial states with off-centre spread.

\subsubsection{Quenching eigenstates of the hermitian Hamiltonian}\label{quenched}
We now turn to the Krylov spread complexity for a  quench from a hermitian to a non-hermitian Hamiltonian at time $t=0$. We consider the system with open boundary conditions through the same study as for the case of periodic boundary conditions. We apply a typical quench protocol where the initial state of the system is taken as one of the eigenstates of the hermitian tight-binding Hamiltonian $H_{\text{TB}}$ given in Eq.~\eqref{eq:tbham_herm}. This state, being an eigenstate of $H_{\text{TB}}$, evolves trivially under the hermitian Hamiltonian as it can only change up to a phase. This trivial evolution for all the times before $t=0$ (starting from $t\rightarrow -\infty$) does not contribute anything to the spread complexity. We then assume that there is a sudden quench at time $t=0$ that shifts $H_{\text{TB}}$ to non-hermitian $H_{\text{eff}}$ given in Eq.~\eqref{eq:tbham_non-herm_open} with open boundary condition. The quench Hamiltonian $H_q$ as a function of time is noted below,
\begin{equation}\label{eq:quench_H}
    H_q = \begin{cases}
    H_{\text{TB}}=H_{\text{TB}}^{\dagger}, & -\infty \leq t < 0 \\
    H_{\text{eff}}\neq H_{\text{eff}}^{\dagger},              & ~~~~~t\geq 0.
\end{cases} \end{equation}

The action of the complex symmetric Hamiltonian constructs the corresponding Hilbert space after $t=0$. Note that in sections \ref{obc} and \ref{pbc}, our initial states were position eigenstates spread over a few sites of the lattice. Therefore, they, not being eigenstates of the tight-binding Hamiltonian or the effective non-hermitian Hamiltonian, would spread under both unitary and non-unitary evolution. However, in this quenching case, the initial state being an eigenstate of the hermitian Hamiltonian evolves non-trivially only after the quench. Therefore, all the evolution that we observe for the state in this case is just due to the measurement process after intervals of $\tau$. 

Let us assume the system is prepared in its single-particle ground state or the single-particle first excited state of the tight-binding hermitian Hamiltonian and evolves nontrivially under the effective non-hermitian Hamiltonian starting at time $t=0$. For such a sudden quenching, we analyse the behaviour of the spread complexity with varying non-hermiticity parameters $\tau$. 
First, we consider the i) the single-particle ground state, and, then ii) the single-particle first excited state of the hermitian Hamiltonian. 

\subsection*{Quenching the single-particle ground state}

\begin{figure}[hbtp]
     \centering
     \begin{subfigure}[b]{0.48\textwidth}
         \centering
         \includegraphics[width=\textwidth]{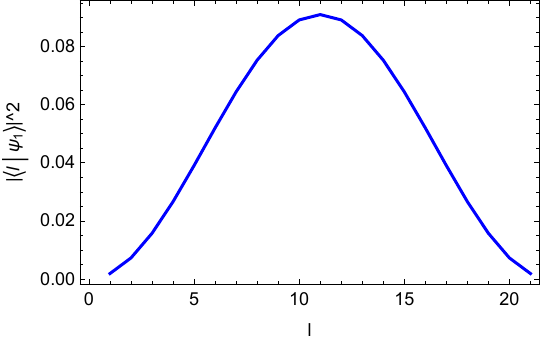}
         \caption{}
         \label{fig:quench_GS}
     \end{subfigure}
     \hfill
     \begin{subfigure}[b]{0.48\textwidth}
         \centering
         \includegraphics[width=\textwidth]{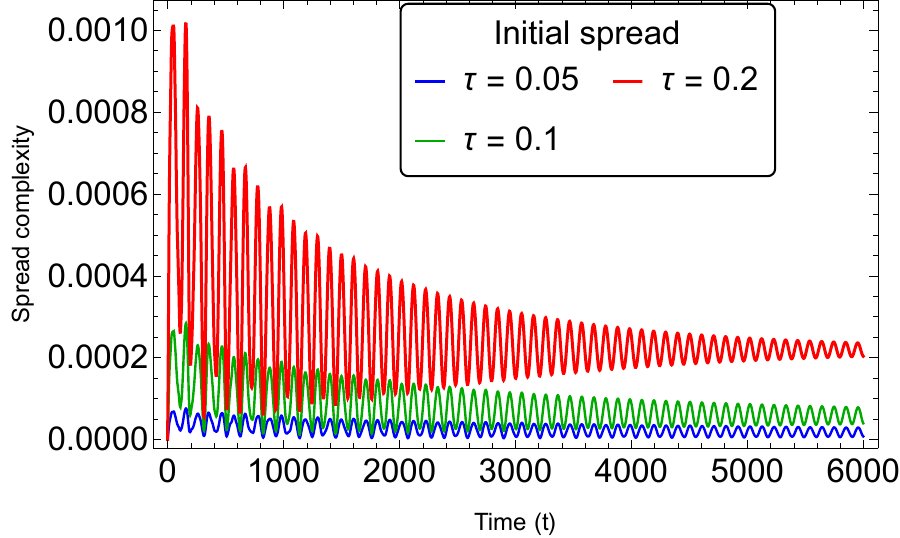}
         \caption{}
         \label{fig:Kry_com_quench_diff_tau}
     \end{subfigure}
        \caption{ (a) Single-particle ground state of tight-binding hermitian Hamiltonian for $N=22$ and (b) Spread complexity for $N=22$ and different values of $\tau=0.05$ (blue), $0.1$ (green), and $0.2$ (red) for the initial state as the single-particle ground state of unperturbed hermitian Hamiltonian. For more frequent measurements, characterized by smaller $\tau$, the initial growth of spread complexity is suppressed and it oscillates around a constant value. }
        \label{fig:comp_entro_quench_gs}
\end{figure}

In this case, we evolve the single-particle ground state of the hermitian Hamiltonian, denoted by $|\psi_{\text{GS}}\rangle$, with non-hermitian $H_{\text{eff}}$. 
\begin{equation}
    |\psi(t=0)\rangle= |\psi_{\text{GS}}\rangle.
\end{equation}
Note that $|\psi_{\text{GS}}\rangle$ has maximum support on the central site of the chain as shown in Figure \ref{fig:quench_GS} with the distribution similar to the steady state of the open boundary conditions (Figure \ref{fig:steady_state_open}).   

 Figure \ref{fig:Kry_com_quench_diff_tau} reveals that when the time gap ($\tau$) between two measurements is large, then there is a notable initial growth in the complexity.
The subsequent decay and saturation structure mirrors previous findings in the open boundary condition scenario. Conversely, for smaller values of $\tau$, the initial growth is relatively insignificant and is comparable to the oscillations in the saturation region. This observation indicates minimal effects of the non-hermitian perturbation on the complexity.

A transition in complexity behaviour is observed based on the $\tau$ parameter of the Hamiltonian. Changing from very small values of $\tau$ of the order of $10^{-2}$ to the values of the order $10^{-1}$ results in a sharp increase in complexity at the initial time followed by a long decay period and a higher value of saturation in Figure \ref{fig:Kry_com_quench_diff_tau}. Now, $\tau$ being the time gap between measurements, this change indicates a shift in the measurement frequency. As we decrease $\tau$ to very small values, it indicates a very high frequency of measurements. In this limit, the complexity oscillates around a constant value without proper growth or decay. 

To understand this, we first recall the probability density of the ground state (Figure \ref{fig:quench_GS}) of the hermitian Hamiltonian expressed in the lattice site position basis, as given by Eq.~\eqref{eq:tbham_herm}. This has a similar profile as the steady states in open boundary conditions (Figure \ref{fig:steady_state_open}). For very small gaps between the measurements, the state does not get enough time to expand and explore the whole Krylov space. The high frequency of the measurements keeps the initial profile of the density matrix almost unchanged. This is the closest limit possible to the quantum Zeno effect in this model \cite{Sudarshan_1977,FACCHI200012, Facchi_2008}, since for $\tau=0$ there is no time evolution at all.  For larger gaps between two consecutive measurements, the initial state spreads further in the Hilbert space before reaching the final steady state. This transition in complexity is a purely measurement-induced phenomenon. It is reminiscent of the measurement-induced phase transition in quantum systems.

\begin{figure}[hbtp]
     \centering
     \begin{subfigure}[b]{0.48\textwidth}
         \centering
         \includegraphics[width=\textwidth]{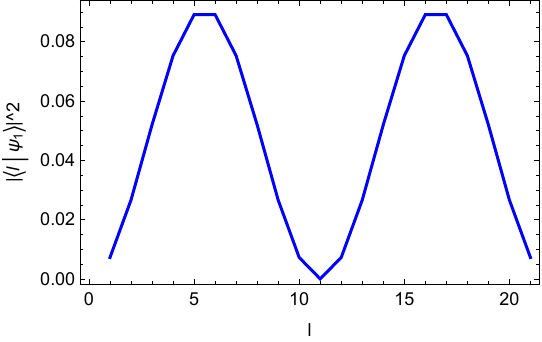}
         \caption{}
         \label{fig:quench_1es}
     \end{subfigure}
     \hfill
     \begin{subfigure}[b]{0.48\textwidth}
         \centering
         \includegraphics[width=\textwidth]{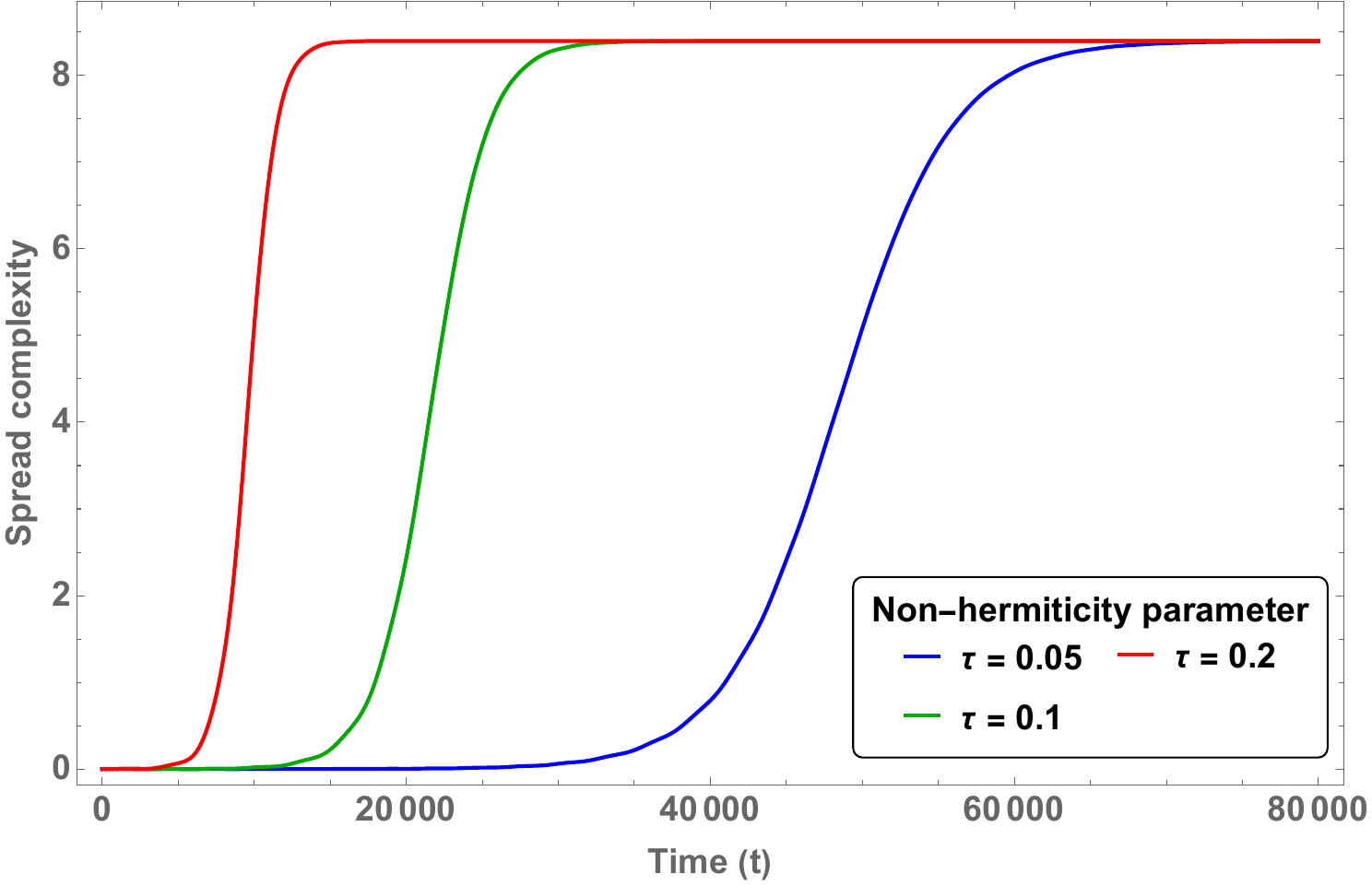}
         \caption{}
         \label{fig:Kry_com_quench_diff_tau_hes}
     \end{subfigure}
         \caption{ (a) Single-particle first excited state of unperturbed hermitian Hamiltonian for $N=22$ and (b) Spread complexity for $N=22$ and different values of $\tau=0.05$ (blue), $0.1$ (green), and $0.2$ (red) for the initial state as the single-particle first excited state eigenstate of unperturbed hermitian Hamiltonian. For smaller $\tau$, the spread complexity starts growing at a later time indicating the Zeno effect.}
        \label{fig:comp_entro_quench_1es}
\end{figure}

Due to the fact that $|\psi_{\text{GS}}\rangle$ as shown in Figure \ref{fig:quench_GS} is already similar to the steady state shown in Figure \ref{fig:steady_state_open} in terms of support on the sites, the overall value of spread complexity is suppressed. For an enhanced quantum Zeno effect, it is thus necessary to consider instead the evolution of the
 single-particle first excited eigenstate of the hermitian Hamiltonian, denoted by $|\psi_{\text{FES}}\rangle$ and shown in Figure \ref{fig:quench_1es}. Since this state has a minimum at the central site, it differs significantly from the final steady state. Thus the quantum Zeno becomes more prominent.

\subsection*{Quenching the single-particle first excited state}

 When the first excited state  $|\psi(t=0)\rangle=|\psi_{\text{FES}}\rangle$, is taken as initial state, it has to explore a significant number of Krylov basis vectors before reaching the steady state. We plot the spread complexity for this choice of initial state in Figure \ref{fig:comp_entro_quench_1es}. 
We notice a clear difference in complexity profile from Figure \ref{fig:Kry_com_quench_diff_tau}. The larger saturation value of the complexity indicates that the steady state is much more complex for the first excited state as compared to the ground state. From the plots for different values of $\tau$, we also clearly notice that the spread complexity starts growing much earlier (red plot) for larger values of $\tau=0.2$. This clearly indicates the measurement-induced phase transition in correlation with the quantum Zeno effect \cite{Sudarshan_1977,FACCHI200012, Facchi_2008}. As we decrease the interval between two consecutive measurements $\tau$, the state takes longer to start evolving. Therefore, we find that when $\tau$ is infinitesimally small, the spread complexity stays at zero for longer. This is precisely equivalent to the statement of the quantum Zeno effect, where a system can not undergo any evolution if projective measurements are made continuously. It is also worth noticing that saturation values of the spread complexity are the same for different values of $\tau$. It indicates that they indeed reach the same steady state after non-unitary time evolution.



\section{Discussion and applications}\label{sec:conclusion}

As we showed, the spread complexity may be used to characterize the non-unitary evolution of  states induced by measurements at regular intervals. We exemplified this by considering the  quantum first passage problem in which regular projective measurements track whether the state has reached the detector site, making the system effectively non-unitary. Our numerical results as given in  section \ref{sec:results} display the effects of non-unitarity based on varying spreads of the initial state and the non-hermitian perturbation parameter $\tau$ that corresponds to the time between two consecutive measurements. We characterize this non-unitary behaviour using the total probability in the Krylov basis, the spread complexity and the spread entropy. In the following, we discuss the main conclusions from our findings.

     \subsection*{Non-unitarity} 
     We implement a time-dependent normalisation for state vectors in the Krylov basis. This normalisation keeps the total probability in the Krylov basis constant at value one with time. As discussed in section \ref{sec:results}, for the  quantum first passage problem (QFPP) the spread complexity and spread entropy behave in a qualitatively similar way as the Krylov complexity for operators in open quantum systems as considered in \cite{Bhattacharya:2023zqt}. The spread complexity in the QFPP grows initially and displays an elongated decay phase before saturating to a constant value (see Figure \ref{fig:kry_com_open}). Note that as discussed in Appendix \ref{appb1}, even for unitary evolution there is a very small decay of the spread complexity after the peak, i.e.~there is an overshooting of the asymptotic value. However, in the case of non-unitary evolution, this decay region persists for a longer period of time. Therefore, the saturation value is significantly suppressed as compared to the peak value.  This decay is caused by the imaginary part of the diagonal coefficients $\alpha_n$ of the tri-diagonal matrix $\Tilde{T}_j$.
     The saturation value of the spread complexity is determined by a steady state of the non-unitary quantum system. This steady state in non-unitary is analogous to thermal states with effectively unitary time evolution. For open boundary conditions, we find that the steady state is universal and unique  for all choices of initial states and non-unitarity parameter $\tau$ (see Figure \ref{fig:steady_state_open}). On the other hand, for  periodic boundary conditions, there is a hierarchy of steady states for different choices of initial state (see Figure \ref{fig:steady_state_periodic}).

    \subsection*{Non-hermiticity parameter $\tau$ } According to Eq.~\eqref{eqtau}, the parameter $\tau$ controls the amount of non-hermiticity. It corresponds to the time between two successive measurements. While the measurements introduce non-unitarity into the problem, states evolve unitarily during the time gap $\tau$ between the two such measurements. Frequent projective measurements impede this evolution. The system's non-unitarity actually spreads during the unitary evolution windows. Therefore, increasing $\tau$ results in a faster decay of the spread complexity and makes it saturate sooner (see Figure \ref{fig:kry_com_diff_tau}). The same saturation value of both complexity and entropy for different values of $\tau$ indicates that given an initial choice of state, the steady state is universal and invariant of $\tau$. A larger value of $\tau$ implies faster equilibration to the steady state. 
    
    For the quench setup, we also find that the saturation value is independent of the non-hermiticity parameter $\tau$, as shown in Figure \ref{fig:comp_entro_quench_1es}. However,  the time when the complexity starts growing increases when the interval $\tau$ between two measurements is decreased. This is reminiscent of the quantum Zeno effect \cite{Sudarshan_1977,FACCHI200012, Facchi_2008}. This effect precisely amounts to  frequent repeated measurements impeding the state evolution. This occurs for $\tau \rightarrow \, 0$. In the Zeno regime, the spread complexity vanishes, as the state does not evolve at all.


    \subsection*{Initial state dependence }
    
   The spread complexity also depends on the choice of the initial state by considering two further conditions, namely i) the spread of the initial state in the position basis and ii) the distance of the initial spread from the location of the detector. For both these cases, our results indicate that the effect of non-hermiticity induced by the measurement occurs at a given distance from the detector after the time necessary for the signal to reach the detector. Therefore, i) increasing the initial spread and ii) decreasing the distance of the initial spread from the detector both make the spread complexity decay and saturate faster to different values (see Figure \ref{fig:kry_com_open}). These different saturation values represent how complicated the steady state is compared to the initial state. For open boundary conditions, in the tight-binding chain the average position of the steady state is located at the centre. For periodic boundary conditions, however, when the initial state is spread close to the detector and does not include the central site of the chain, there is no localised steady state, as it keeps oscillating over the entire tight-binding chain. Both spread complexity and entropy respect the hierarchy of the saturation value in a coherent manner.

    \subsection*{Behaviour of Lanczos coefficients } 
    
    We find evidence that the behaviour of Krylov complexity or entropy is not entirely dictated by the Lanczos ascent and descent behaviours. Previous papers \cite{Parker:2018yvk, Balasubramanian:2022tpr, Erdmenger:2023shk} have stressed the different behaviour of Lanczos coefficients distinguishing the integrable or chaotic nature of the theory. In particular, in the operator growth hypothesis, the growth and saturation behaviour of complexity was correlated to the growth of the Lanczos coefficients \cite{Parker:2018yvk, Alishahiha_2023}.   Here we find for non-unitary systems that the spread complexity and entropy still show the characteristic time evolution of growth-peak-decay-saturation form, even if the coefficients are oscillatory. This is discussed in Appendix \ref{appb2}. As long as the operator generating the evolution is written in a tridiagonal form, the complexity shows regimes of growth, decay and saturation. We note that even for a tight-binding chain without measurements, corresponding to taking $\tau = 0$ and thus a limiting unitary case, spread complexity and entropy grow and saturate though the Lanczos coefficients are oscillatory (refer to Appendix \ref{appb1}).

    \subsection*{Spread entropy }
    
    The spread entropy in the Krylov basis saturates to a plateau after initial growth (as shown in Figures \ref{fig:kry_ent_open} and \ref{fig:kry_ent_diff_spd_per}). For non-unitary evolution with open and periodic boundary conditions in QFPP, the approximate relation between spread entropy ($S_S$)  and complexity ($C_S$), conjectured previously for unitary systems in \cite{Balasubramanian:2022tpr}  ($C_S\approx e^{S_{S}}$), persists. In the case of periodic boundary conditions, we observe high oscillations in the saturation phase of spread entropy, showing effective unitary behaviour when the initial states are spread near the detector and do not include the central site (see Figure \ref{fig:comp_entro_periodic}). 
    
    \subsection*{Distance from detector }
    
    The distance of the spread of the initial state from the detector plays an important role in the complexity dynamics.  If the detector is placed at one boundary ($N$th site) for a finite number of sites ($N$ sites), the nature of the plots is symmetric for when i) the initial spread is between $\left(n_2 -n_1\right)$ sites and ii) the initial spread is between $\left(N-(n_2 -n_1) \right)$  sites. This symmetry results from the expression of the survival amplitude \cite{Dhar2013QuantumTO, Dhar_2015} mentioned in Eqs. ~\eqref{survival1} and \eqref{survival2}. Since the Lanczos coefficients can be obtained from the moments of survival amplitude \cite{VM} as shown in Appendices \ref{lanczosfsurv} and \ref{lanczosfsurv2}, our work shows that this symmetry is reflected in the behaviour of spread complexity and entropy as well.

    \subsection*{Periodic BC and effective unitarity }QFPP with periodic boundary condition requires additional attention as discussed in section \ref{pbc}. This may also be seen by considering the \textit{survival probability} instead of building the Krylov basis using the Lanczos algorithm. Previous studies in \cite{Dhar_2015} showed that in this case, the survival probability, defined in appendix \ref{lanczosfsurv} in Eq.~\eqref{return_prob}\footnote{Note that the definition of survival probability is slightly different in the paper \cite{Dhar_2015}. There it is denoted as $\langle \psi(t)|\psi(t)\rangle$.} decays to zero only if the initial spread of the state for a $N$ site Hamiltonian is symmetric with respect to the $\frac{N}{2}$-th site. Else, it decays to a non-zero constant value and remains saturated. This saturation value is $0.5$ in cases when the initial spread does not contain the central site. We find the exact same behaviour in the time-dependent spread probability (Appendix \ref{app:K_probability}). Therefore, our results in this setup indicate a correspondence between the survival probability and the time-dependent Krylov spread probability\footnote{This is to be expected since we already know that the Lanczos elements can be equivalently found directly from the moments of the survival probability.}. We also find that after the time-dependent probability becomes constant, the spread complexity and entropy also behave similarly to unitary setups (Appendix \ref{appb1}), i.e., the complexity and entropy both saturate after decaying for a lesser amount of time. Therefore, in this case, we observe an effective unitary nature of the system after the survival probability saturates to non-zero constant values. The Krylov basis can capture this effect very elegantly. Of course, for other cases when the initial spread contains the central site, the behaviour is like the usual non-unitary cases mentioned earlier, as the time-dependent probability decays all the way to zero.

    \subsection*{Quench } In section \ref{quenched}, our initial state is the single-particle ground state of the actual hermitian Hamiltonian, and therefore, it evolves under the unitary evolution with a phase factor only. This phase factor does not contribute to the spread complexity. This means that if we evolve this state unitarily for time $t_1$, the complexity remains at zero. At time say, $t_1$, we introduce the detector and start measurements at regular intervals $\tau$ given by Eq.~\ref{eq:quench_H}. Now, the ground state of the hermitian Hamiltonian starts evolving nontrivially since it is not an eigenstate of the effective non-hermitian Hamiltonian. As a result, complexity starts showing non-unitary evolution. This can be understood as a unitary-to-non-unitary quench taking place at time $t_1$  that is time zero in our plots in section \ref{quenched}. Spread complexity is a good probe of this transition since it changes behaviour at the point of the transition time. By modulating the measurement frequency, we discern a notable shift in the behaviour of spread complexity. For smaller values of the time gap between measurements of the order $10^{-2}$, we observe that the complexity oscillates around a constant value. Conversely, as we increase the value of $\tau$ to the order $10^{-1}$, the complexity exhibits an initial phase of rapid growth, succeeded by a prolonged decaying region and eventual saturation — a characteristic signature of spread complexity for non-unitary evolution. Consequently, we infer that more frequent measurements restrict the system to evolve for only shorter duration post-measurement, thereby constraining the growth of complexity. This is reminiscent of the measurement-induced phase transition phenomena \cite{measure1, measure2} in the context of complexity growth. 
    
    The transition in the complexity behaviour becomes more prominent for the quench of the single-particle first excited state. Even though the initial spread of this state on the lattice sites is much different from the steady state (see Figure \ref{fig:quench_1es} and Figure \ref{fig:steady_state_open}), the state does not go through any nontrivial evolution after the quench for a long time if the measurement frequency is very high. This is because of the quantum Zeno effect taking place that suggests frequent measurements impede the time evolution and spreading of a quantum state. Finally, after a very long time that depends on the value of the $\tau$ parameter, complexity starts growing rapidly, and it saturates to a constant value when the system reaches a steady state. In this case, contrary to all the previous cases, the evolution directly leads to a steady state without much oscillation. 

    To conclude further, our results show that the spread complexity and entropy in the Krylov basis can act as very good probes for different phenomena during a non-unitary evolution ranging from the varying non-hermiticity to effective unitarity in periodic boundary conditions and unitary-nonunitary transition through quench. This article successfully generalizes the concept of spread in the Krylov basis to non-unitary setups. Our work further strengthens the idea of implementing the modified approach of bi-Lanczos and complex symmetric Lanczos for non-unitary cases as advocated in \cite{Bhattacharya:2023zqt}. We find an agreement between the non-unitary Lanczos coefficients to the ones derived directly from the non-unitary survival amplitude for the first time. It shows that the survival amplitude is a more fundamental way to compute Lanczos coefficients in the sense that the moment recursion method remains unchanged for a non-unitary setup, whereas the construction through the actual Lanczos algorithm needs to be modified.

    \subsection*{Applications of Tight-Binding model}
    Since we have studied the effect of measurement using a tight-binding Hamiltonian, our results can be useful in multiple physical situations where a tight-binding Hamiltonian is used in model building. A couple of examples of very active areas of research both from theory and experiment points of view are, 

    i) electron transport in one-dimensional systems \cite{Karamlou_2022},    

    ii) the behaviour of electrons confined to one-dimensional structures, such as quantum wells and quantum dots \cite{Sanderson2009QuantumDG}.

    For these cases, the diagonal and off-diagonal entries of the Hamiltonian can be understood as the orbital energies of an atomic orbital  and the hopping amplitudes of an electron to jump to a different overlapping orbital of another atom, respectively. These help to determine the electronic transport properties of nanostructures. Our results indicate the simple fact that  measurements of these diagonal and hopping amplitudes are expected to give imaginary results in general, i.e.~to have a complex phase. This imaginary nature will reflect the non-hermiticity of the Hamiltonian due to the effect of measurements. Therefore, the effective non-hermitian Hamiltonian ($H_{\text{eff}}$), under which the system  evolves when measurements are made, can be written from those observed imaginary hopping amplitudes. The methods introduced in this paper can be used to determine the exact time dynamics by performing the Krylov basis analysis and studying the spread complexity for different electronic configurations. We hope that this may be useful in further characterization of these nanostructures.

    \subsection*{Future directions}
    The techniques developed in this work are expected to also be useful in the studies of open system dynamics in the Schrödinger picture, where a mixed-state density matrix evolves with a non-hermitian Lindbladian, similar in spirit to the study of \cite{Bhattacharya:2022gbz, Bhattacharya:2023zqt} for operators.  These studies of open system dynamics are also interesting from the perspective of black hole evolution in a thermal environment. This can be extended to the Schwinger-Keldysh path integrals study open quantum field theories building on \cite{Baidya:2017eho, Geracie:2017uku}. It will also be interesting to perform a similar study for the two-dimensional tight-binding Hamiltonian, which is expected to provide information about the time dynamics of electronic transport in 2D graphene-like materials. Another direction that we are pursuing presently, and hope to report soon, is to study PT-symmetric Hamiltonians and how spread complexity and entropy probe the PT-symmetry breaking in such systems \cite{spread2upcoming}. 
    
    A final future direction is to perform a similar study in holography.  There are recent studies concerning measurements in the SYK model as well as holographic measurements in JT gravity \cite{Jian:2021hve,Antonini:2022lmg}. It will be very interesting to check whether there is an exact match of Krylov spread complexities under holographic measurements, extending the work on holographic Krylov complexity of \cite{Rabinovici:2023yex}. 
    
\raggedbottom
\section*{Acknowledgements}
The authors would like to thank Vijay Balasubramanian, Souvik Banerjee, Pablo Basteiro, Pawel Caputa, Mario Flory, Giuseppe Di Giulio, Taishi Kawamoto, René Meyer, Sara Murciano, Pratik Nandy, Pingal Pratyush Nath, Subir Sachdev, Himanshu Sahu, Aninda Sinha, Zhuo-Yu Xian, Kunal Pal, Kuntal Pal, and Diptarka Das for useful discussions and comments on related topics at various stages. The work of A.B. is supported by the Polish National Science Centre (NCN) grant 2021/42/E/ST2/00234. R.N.D.~and J.E.~are supported by Germany's Excellence Strategy through the W\"urzburg‐Dresden Cluster of Excellence on Complexity and Topology in Quantum Matter ‐ ct.qmat (EXC 2147, project‐id 390858490),
and  by the Deutsche Forschungsgemeinschaft (DFG) 
through the Collaborative Research centre \enquote{ToCoTronics}, Project-ID 258499086—SFB 1170. R.N.D.~further acknowledges the support by the Deutscher Akademischer Austauschdienst (DAAD, German Academic Exchange Service) through the funding programme, \enquote{Research Grants - Doctoral Programmes in Germany, 2021/22 (57552340)}. This research was also supported in part by Perimeter Institute for Theoretical Physics. Research at Perimeter Institute is supported by the Government of Canada through the Department of Innovation, Science and Economic Development and by the Province of Ontario through the Ministry of Research, Innovation and Science. B.D.~acknowledges MHRD, India for Research Fellowship. B.D.~would also like to acknowledge the support provided by Max Planck Partner Group grant MAXPLA/PHY/2018577. B.D. would further like to acknowledge the support provided by the MATRICS grant SERB/PHY/2020334.

\appendix

\section{ Time-dependent probability for non-unitary evolution}\label{app:K_probability}

Here we discuss details of the time-dependent probability in the Krylov basis before implementing the new normalisation described in section \ref{nhnewnorm}. This probability captures the rate of decay of the state through the detector site due to repeated measurements. 

\subsection{ Open boundary conditions}

We begin with the results for open boundary conditions with $N$ sites, with the detector being present at the $N^{th}$ site only. This is characterized by the effective Hamiltonian given in Eq.~\eqref{eq:tbham_non-herm_open}. We first examine the decay of the total probability of the state in the Krylov space. After we spread the time-evolved state on the Krylov basis (using the complex symmetric Lanczos construction), the final state expansion looks like the following,
\begin{equation}
|\psi(t)\rangle=\sum_{n=1}^D\psi_n(t)|K_n\rangle.
\end{equation}
First, we investigate the temporal behaviour of the time-dependent probability in the Krylov space $\sum_{i=1}^D|\psi_n(t)|^2$. Note that this is different from the notion of probability we discussed in Eq.~\eqref{nhdensity}, where it was renormalised to $1$ for all times. The time-dependent probability provides us with an understanding of how the information is lost slowly from the system to the environment due to repeated measurements.

\subsubsection{Probability decay} 
\begin{figure}[H]
     \centering
     \begin{subfigure}[b]{0.49\textwidth}
         \centering
    \includegraphics[width=\textwidth]{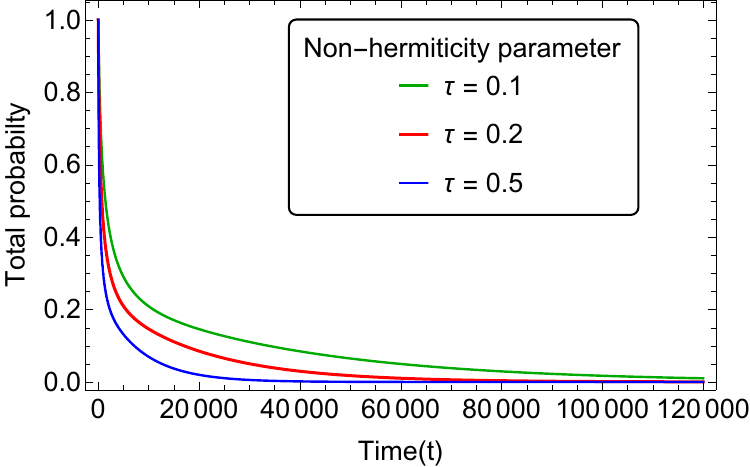}
         \caption{Decay of time-dependent probability with time for $N=40$ and the initial spread of $18\leq l \leq 22$ for different values of non-hermitian perturbation parameter $\tau=0.1$ (green), $0.2$ (red), and $0.5$ (blue).}
         \label{fig:prob_diff_tau_open}
     \end{subfigure}
     \hfill
     \begin{subfigure}[b]{0.49\textwidth}
         \centering
         \includegraphics[width=\textwidth]{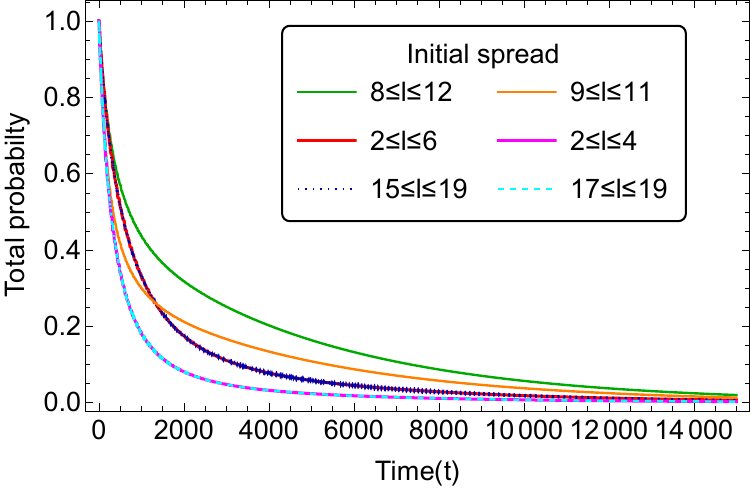}
         \caption{Decay of time-dependent probability with time for $N=20$ and $\tau=0.1$ for different spreads of the initial state $8\leq l \leq 12$ (green), $9\leq l \leq 11$ (orange), $2\leq l \leq 6$ (red), $2\leq l \leq 4$ (pink), $15\leq l \leq 19$ (dotted blue), and $17\leq l \leq 19$ (dotted cyan).}\label{fig:prob_dif_spd_opn}
     \end{subfigure}
        \caption{Decay of time-dependent probability with open boundary conditions for different values of the non-hermiticity parameter and initial spread. The probability decays faster for larger $\tau$ and off-centre spread of initial state.  }
        \label{fig:prob_opn1}
\end{figure}

Due to the non-unitary nature of the time evolution, the preservation of the time-dependent probability is not possible. The decay of time-dependent probability to zero exemplifies the dissipative characteristics induced by the measurements in the system. Notably, in Figure \ref{fig:prob_diff_tau_open}, as the value of $\tau$ increases from $0.1$ (green) to $0.5$ (blue) through $0.2$ (red) for an initial spread in the sites $18\leq l\leq 22$ for a chain of length $40$, the decay rate becomes faster, indicating a faster dissipative process or more rapid detection of the state at the detector. This indicates that increasing the non-hermiticity parameter results in a stronger interaction between the system and the rest of the universe through the detection measurement.

The decay rate of time-dependent probability also depends on the initial state configuration. In Figure \ref{fig:prob_dif_spd_opn}, we show the time dependence of  this probability for the different spreading of the initial state over a chain with total sites $N =20$, keeping the non-hermiticity parameter fixed. Rates of decay for the total probability depend on how wide the initial state is spread over the sites. In particular, a wider spread of the initial state corresponds to a slower decay rate. For example, in Figure \ref{fig:prob_dif_spd_opn}, the total probability corresponding to the initial spread $8\leq l\leq 12$ (green line) decays slower than the total probability corresponding to the initial spread $9\leq l\leq 11$ (orange line).

We also observe that the decay rates of total probability depend on the location of the initial state. Total probability decays slowly with time for the state initially spread around the midpoint of the system. In Figure \ref{fig:prob_dif_spd_opn}, for instance, the total probability for initial spreads of $2\leq l\leq 6$ (red) decays more quickly than the total probability for initial spreads of $8 \leq l \leq 12 $ (green). Consequently, as the initial states deviate from the centre of the system, decay rates of total probability increase.

Furthermore, despite the detector being situated solely at the end of the chain, the decay rates exhibit symmetry with respect to the distance from the centre rather than the distance from the detector. As from Figure \ref{fig:prob_dif_spd_opn}, the initial state corresponding to the red line $( 2 \leq l \leq 6 )$ and blue dotted line $( 15 \leq l \leq 19)$. Also, the pink line $( 2 \leq l \leq 4 )$ and cyan dotted line $( 17 \leq l \leq 19)$ spread over symmetrically from the centre of the chain. The total probability corresponding to all the initial states with a symmetric spread around the centre decays at the same rate.
This behaviour aligns well with the symmetrical nature of the survival probability in this system \cite{Dhar_2015}. 

In the next subsection, we discuss an alternative approach to obtaining the same set of complex Lanczos coefficients. 

\subsubsection{Lanczos coefficients from survival amplitude}\label{lanczosfsurv}
Like the usual case of hermitian Hamiltonians, the Lanczos coefficients can also be calculated by using the moments obtained from the survival amplitude \cite{VM}, 
\begin{equation}
    S(t)=\langle\psi(t)|\psi(0)\rangle. \label{return_prob}
\end{equation}
The initial state spread in the position basis can be expressed as, 
\begin{equation}
    |\psi(t=0)\rangle=\frac{1}{\sqrt{|l_2-l_1+1|}}\sum_{l=l_1}^{l_2}|l\rangle.
    \label{eq:state}
\end{equation} 
Following the technique described in \cite{Dhar_2015}, using the $V_\text{eff}$ as perturbation and calculating till the first order in the perturbation, we obtain the general analytical form of the survival amplitude for the open boundary condition as,
\begin{equation}\label{survival1}
    S(t)= \frac{1}{l_2-l_1+1}\sum _{s=1}^N \sum _{l=l_1}^{l_2} \sum _{k=l_1}^{l_2} \frac{2}{N} \sin \left(\frac{\pi  k s}{N}\right) \sin \left(\frac{\pi  l s}{N}\right) \exp \left[-t \gamma_s\right]
\end{equation}
where $\gamma_s=\left(\frac{\tau}{N}  \sin ^2\left(\frac{\pi  s}{N}\right)+2 i \cos \left(\frac{\pi  s}{N}\right)\right)$. From this, it is straightforward to get any the higher moment as,
\begin{equation}
    \mu_n=\frac{d^n}{dt^n}S(t)|_{t=0}=\frac{1}{l_2-l_1+1}\sum _{s=1}^N \sum _{l=l_1}^{l_2} \sum _{k=l_1}^{l_2} \frac{2(-\gamma_s)^n}{N} \sin \left(\frac{\pi  k s}{N}\right) \sin \left(\frac{\pi  l s}{N}\right).
\end{equation}
The Lanczos coefficients can be obtained from these moments recursively \cite{VM}. The exact match between the Lanczos coefficients obtained from these moments and those from the complex symmetric Lanczos algorithm can be easily verified numerically. We have checked this all along and found agreement in each of the cases we studied.  Therefore, it is evident that while the Lanczos construction needs to be modified for non-hermitian Hamiltonians in order to obtain meaningful Lanczos coefficients, the moment recursion method from survival amplitude is unaffected. 

\subsection{Periodic boundary conditions}
Next, we look at the system with periodic boundary conditions, when the evolution is governed by the effective Hamiltonian given as Eq.~\eqref{eq:tbham_non-herm_periodic}. In this case, we discuss two main features. Firstly, the Krylov probabilities have an interesting feature with respect to the spread of the initial states, namely an effective unitarity. Secondly, we discuss the implications of this effective unitarity at late times on the complexity and entropy. We only report the results for varying initial states as this is the case where it is significantly different from the open boundary conditions.

\subsubsection{Time-dependent probability}
\begin{figure}[htbp]
	\centering
	\includegraphics[width=\textwidth]{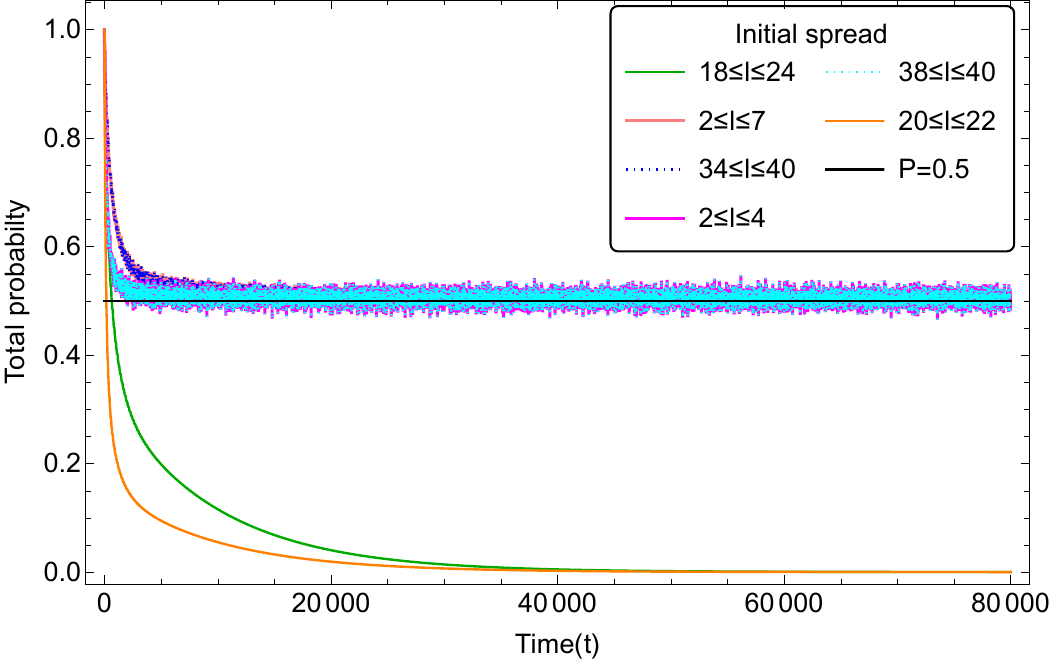}
	\caption{Time variation of total probability with periodic boundary conditions for different spreads of the initial state keeping fixed $N= 40$ and $\tau=0.1$. The black straight line marks the $0.5$ constant line. The time-dependent probability for off-centre initial spread stops decaying earlier and saturates at a constant value characterizing an effective unitarity.}
	\label{fig:Toal_prob_diff_spd_periodic}
\end{figure}
We discussed earlier in section \ref{sec:first-passage} that the decay of the survival probability saturates at $0.5$ when the initial spread of the state does not include the middle site $(N/2)$ at all, i.e.~the spread is in one half of the chain. We also find similar behaviour of the total Krylov probability. Figure \ref{fig:Toal_prob_diff_spd_periodic} shows the time dependence of total probability for different spreading of the initial state with periodic boundary conditions for a fixed number of total sites $N = 42$ and non-hermiticity parameter $\tau = 0.1$. The total probability associated with the initial state of spreading $18 \leq l \leq 24$ (green line) and $20 \leq l \leq 22$ (orange line) decays to zero. On the other hand, we notice the saturation at $0.5$ of the total probability corresponding to the initial state spread over $2 \leq l \leq 4$ (Magenta line), $38 \leq l \leq 40$ (dotted cyan line), $34 \leq l \leq 40$ (dotted blue line), and $2 \leq l \leq 7$ (pink line). This is due to a symmetry that decreases the Krylov space dimension by half when the central site of the chain is not included in the initial state in a symmetric fashion.  If the initial state is spread symmetrically around the central site, it has equal access to both halves; hence, the symmetry does not manifest in the plots. Thus the probability decays completely to zero at late times. However, suppose the initial state is distributed non-symmetrically around the centre of the chain. In that case, the access is divided non-symmetrically into the two halves of the symmetry, and we observe the probability to saturate still but to values lower than $0.5$. This observation is in line with the survival probability property mentioned in the previous works \cite{Dhar_2015, Dhar2013QuantumTO}.

On the other hand, decreasing the non-hermiticity parameter and increasing the spread of the initial state, in this case, results in a slower rate of decay, as we observed in the case of the open boundary conditions.

\subsubsection{Lanczos coefficients from survival amplitude}\label{lanczosfsurv2}
Finally, to complete this section, we again ensure that all of these results can be reproduced, starting from the survival amplitude and computing its moments. If we take an initial state like Eq.~\eqref{eq:state} into consideration, then the expression for the survival amplitude for periodic boundary conditions has the following analytic form as,
 \begin{align}\label{survival2}
S(t)= \frac{2}{N(l_2-l_1+1)} \sum _{l=l_1}^{l_2} \sum _{k=l_1}^{l_2} \Bigg( \sum _{s=0}^{\frac{N}{2}-1} \Phi_{2 s+1}(l) \Phi_{2 s+1}(l) \exp\left(-\gamma_{2s+1}t\right)  \nonumber\\  +\sum _{s=1}^{\frac{N}{2}-1}  \Phi_{2 s}(k) \Phi_{2 s}(l) \exp\left(-2 i t\cos \left(\frac{2 \pi  s}{N}\right)\right) \Bigg).
\end{align}
 Using this, all the higher moments can be written in general as,
 \begin{align}
\mu_n= \frac{2}{N(l_2-l_1+1)} \sum _{l=l_1}^{l_2} \sum _{k=l_1}^{l_2}\Bigg(\sum _{s=0}^{\frac{N}{2}-1} \Phi_{2 s+1}(l) \Phi_{2 s+1}(l) \left(-\gamma_{2s+1}\right)^n \nonumber \\+\sum _{s=1}^{\frac{N}{2}-1}  \Phi_{2 s}(k) \Phi_{2 s}(l) \left(-2 i \cos \left(\frac{2 \pi  s}{N}\right)\right)^n  \Bigg) ,
\end{align}
where we denote $\sin \left(\frac{\pi  k s}{N}\right)$ as $\Phi_s(l)$ to make the equation more compact. From here, the Lanczos coefficients can be obtained in an iterative manner \cite{VM}.

\section{Unitary spread complexity and non-unitary  Lanczos coefficients}
\begin{figure}[hbtp]
	\centering
	\begin{subfigure}[b]{0.48\textwidth}
		\centering
		\includegraphics[width=\textwidth]{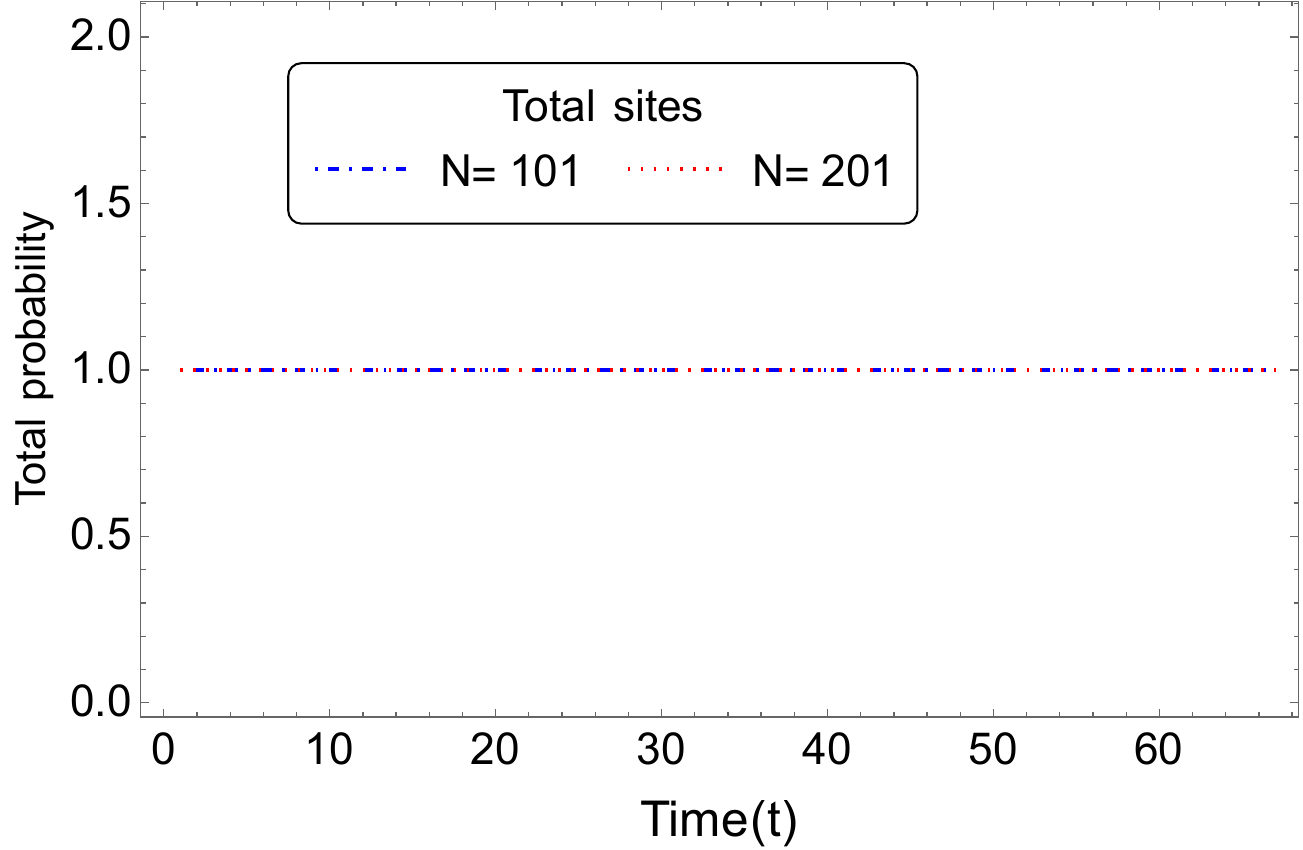}
		\caption{Total probability for unitary evolution remains unchanged with time.}
		\label{fig:probability-unitary}
	\end{subfigure}
	\hfill
	\begin{subfigure}[b]{0.49\textwidth}
		\centering
		\includegraphics[width=\textwidth]{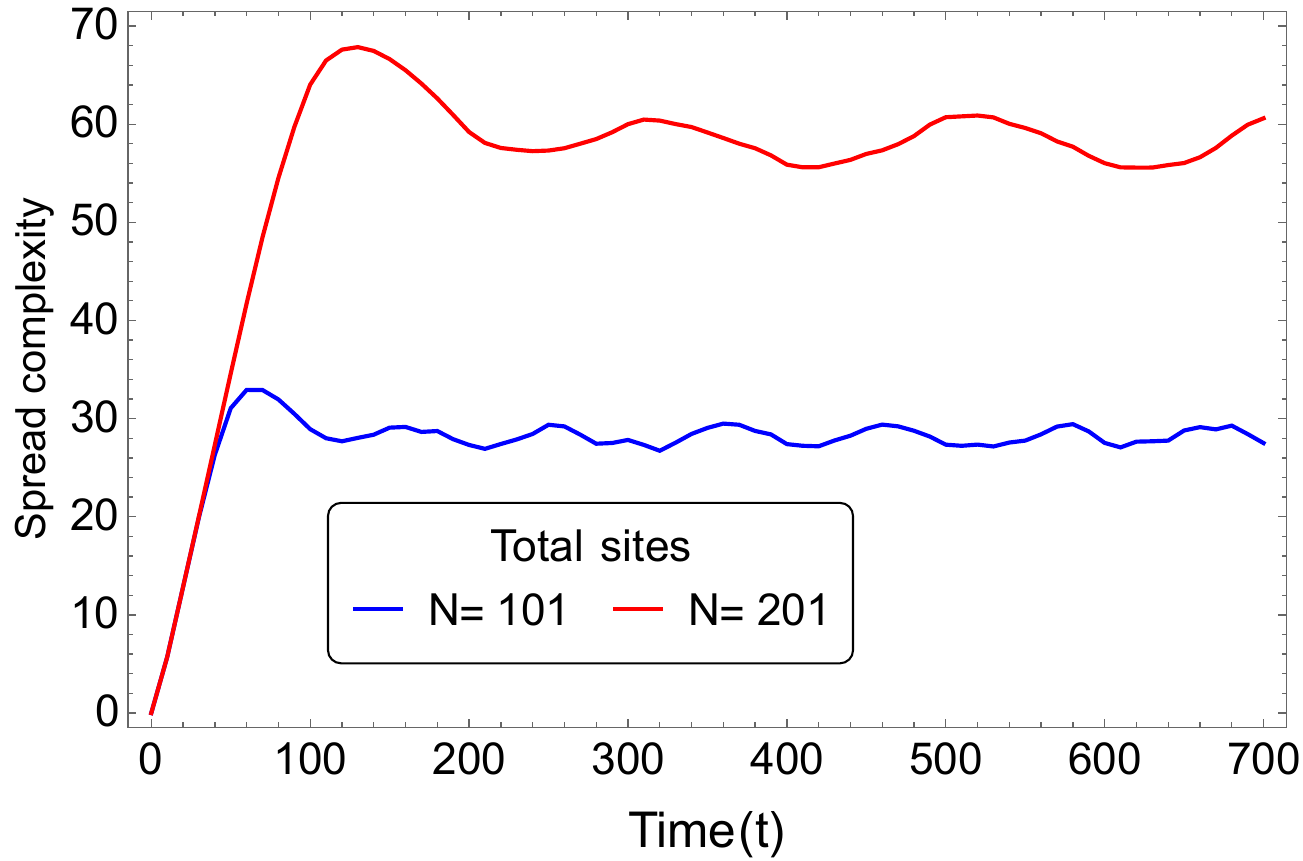}
		\caption{Time evolution of spread complexity for $N = 201$ (red), and $N =101$ (blue).}
		\label{fig:spread-unitary}
	\end{subfigure}
	\begin{subfigure}[b]{0.49\textwidth}
		\centering
		\includegraphics[width=\textwidth]{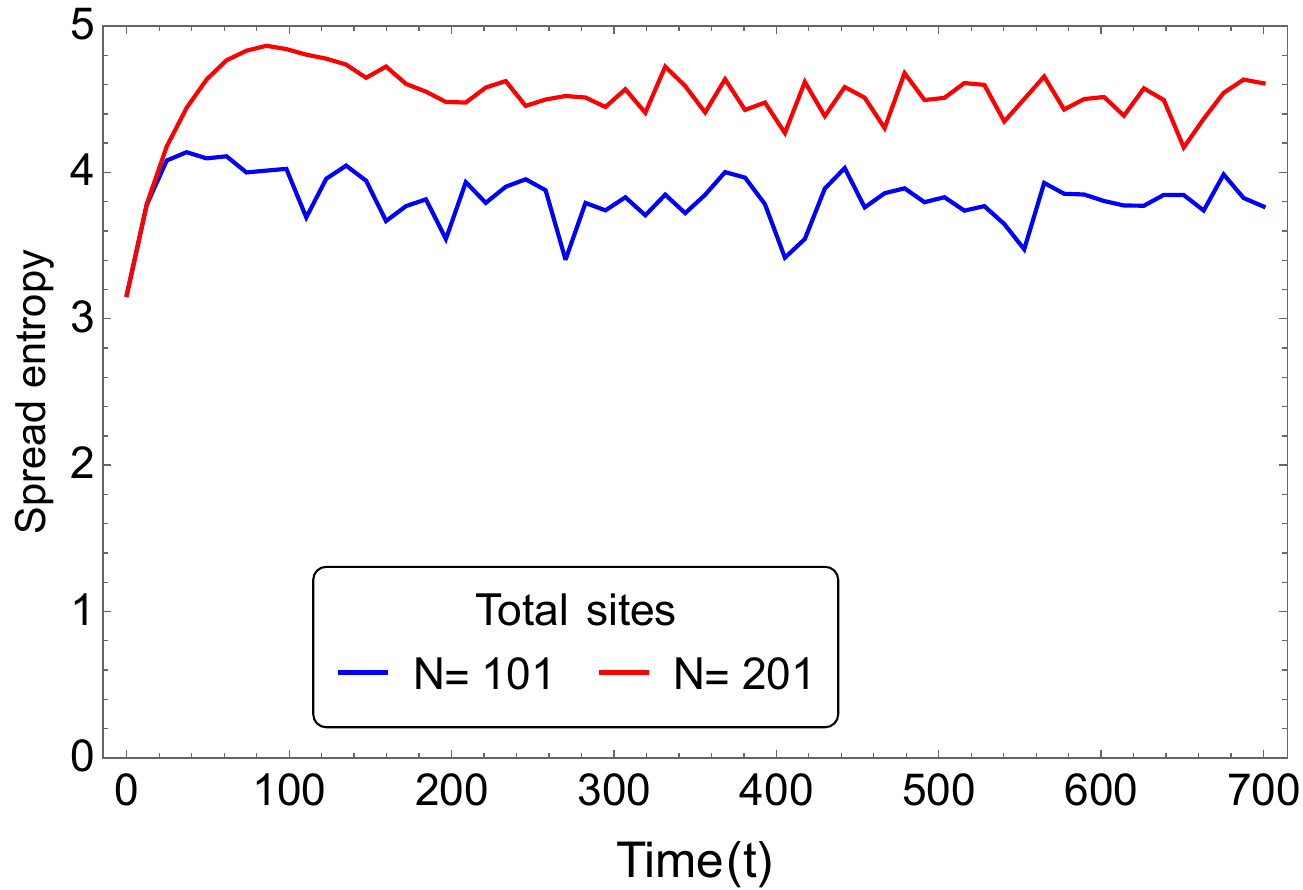}
		\caption{Time evolution of spread entropy for $N = 201$ (red), and $N =101$ (blue).}
		\label{fig:entropy-unitary}
	\end{subfigure}
	\caption{Total probability, spread complexity, and spread entropy for unitary evolution for different numbers of total sites, starting with an initial state spreading over the centre of the chain. The complexity and entropy show growth and saturation at a higher value for larger system size $N$.}
	\label{fig:unitary}
\end{figure}
To illustrate the crucial differences in the behaviour of spread complexity and entropy for non-unitary evolution and unitary evolution, we here provide a numerical analysis for unitary evolution. 

\subsection{Unitary spread in Krylov basis} \label{appb1}
In this section, we provide plots of spread complexity for unitary evolution under the tight-binding Hamiltonian given in Eq.~\eqref{eq:tbham_herm}. The Hamiltonian is very simple, and of the following form,
\begin{equation}
    H=\begin{pmatrix} 0 &-1&&&&0\\-1&  0& -1&&&\\&-1&0&\ddots &&\\&&\ddots &\ddots &-1&\\&&&-1&0&
    \ddots\\0&&&&\ddots&\ddots\\\end{pmatrix}_{N\times N},
\end{equation}
where $N$ is the number of sites. We take $N=201$, $98\leq l\leq 102$ (red) and $N=101$, $48\leq l\leq 52$ (blue) and plot the total probability, spread complexity and spread entropy in Figure \ref{fig:unitary}.

\subsubsection{Probability and complexity}
As expected, Figure \ref{fig:probability-unitary} shows that the probability is conserved throughout. This ensures that the evolution is indeed unitary. The spread complexity saturates to a plateau after initial linear growth (see Figure \ref{fig:spread-unitary}). Therefore the oscillating decay part of spread complexity in the cases studied in the main text is attributed to the non-hermiticity of the effective Hamiltonian, due to which the evolution becomes non-unitary. We also observe that as we increase the system size, the initial growth and the saturation value increases, which is in agreement with previous literature on the subject \cite{Balasubramanian:2022dnj, Parker:2018yvk}.

\subsubsection{Krylov spread entropy} The spread entropy also attains a plateau after the initial logarithmic growth in the unitary case, as shown in Figure \ref{fig:entropy-unitary} and the saturation value increases with increasing Hilbert space dimensions. In a similar spirit to spread complexity, we are therefore able to identify the indefinite growth of the Krylov spread entropy discussed in the main text to be a consequence of the non-unitary evolution.

\subsubsection{Lanczos coefficients for unitary evolution}
\begin{figure}[hbtp]
	\centering
	\begin{subfigure}[b]{0.48\textwidth}
		\centering
		\includegraphics[width=\textwidth]{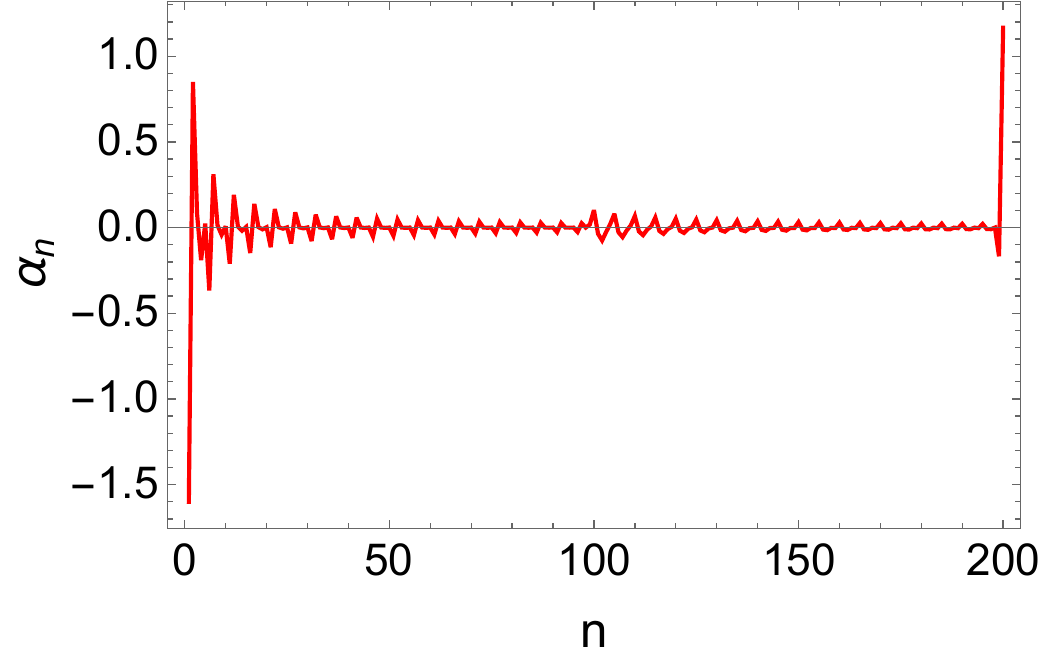}
		\caption{$\alpha_n$ for unitary evolution.}
		\label{fig:an-unitary}
	\end{subfigure}
	\hfill
	\begin{subfigure}[b]{0.49\textwidth}
		\centering
		\includegraphics[width=\textwidth]{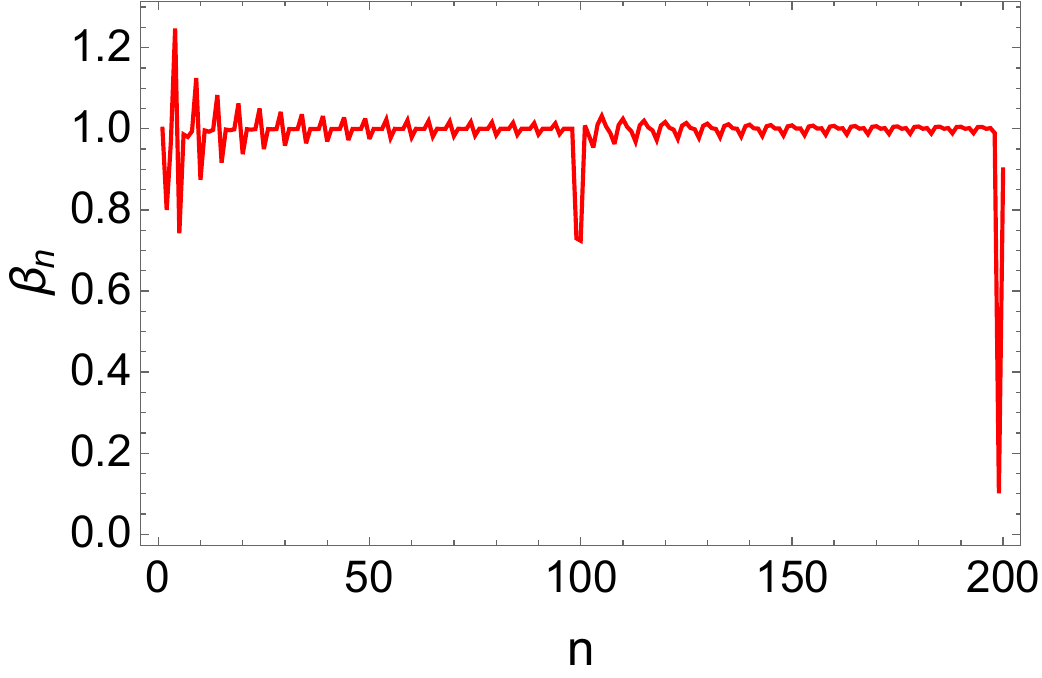}
		\caption{$\beta_n$ for unitary evolution.}
		\label{fig:bn-unitary}
	\end{subfigure}
	\caption{Behaviour of $\alpha_n$ and $\beta_n$ for unitary evolution. We choose $N=201$ and spreading of the initial state in $98\leq l\leq 102$. There is no Lanczos ascent or descent behaviour in the Lanczos coefficients, implying that this is not required for the typical time dependence of complexity and entropy.}
	\label{fig:anbn-unitary}
\end{figure}
In Figure \ref{fig:anbn-unitary}, we plot the Lanczos coefficients for  unitary evolution of the tight-binding Hamiltonian with $N=201$ and initial state to be spread in $98\leq l\leq 102$. It is instructive to observe that the coefficients do not show any distinct ascent or descent, apart from some peaks, and decay very near the boundary and for most of the time, the behaviour is oscillatory. This implies that although the integrable or chaotic nature of a system can be a result of particular scaling of the ascent and descent of $\beta_n$ and $\alpha_n$ coefficients in some cases, this ascending or descending behaviour is not necessary to give rise to the expected typical spread complexity and entropy profiles.

\subsection{Lanczos coefficients for the non-hermitian Hamiltonian}\label{appb2}
\begin{figure}[hptb]
	\centering
	\begin{subfigure}[b]{0.48\textwidth}
		\centering
		\includegraphics[width=\textwidth]{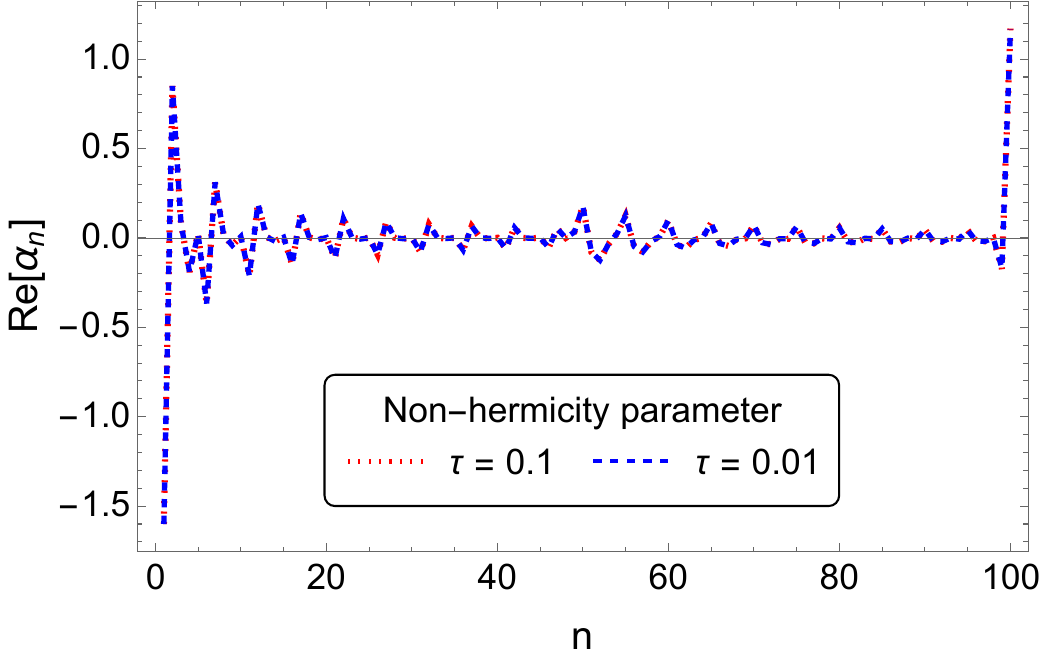}
		\caption{Real parts of $\alpha_n$ for $\tau = 0.1$(dotted red) and $\tau = 0.01 $(dashed blue).}
		\label{fig:rean-open-mid}
	\end{subfigure}
	\hfill
	\begin{subfigure}[b]{0.48\textwidth}
		\centering
		\includegraphics[width=\textwidth]{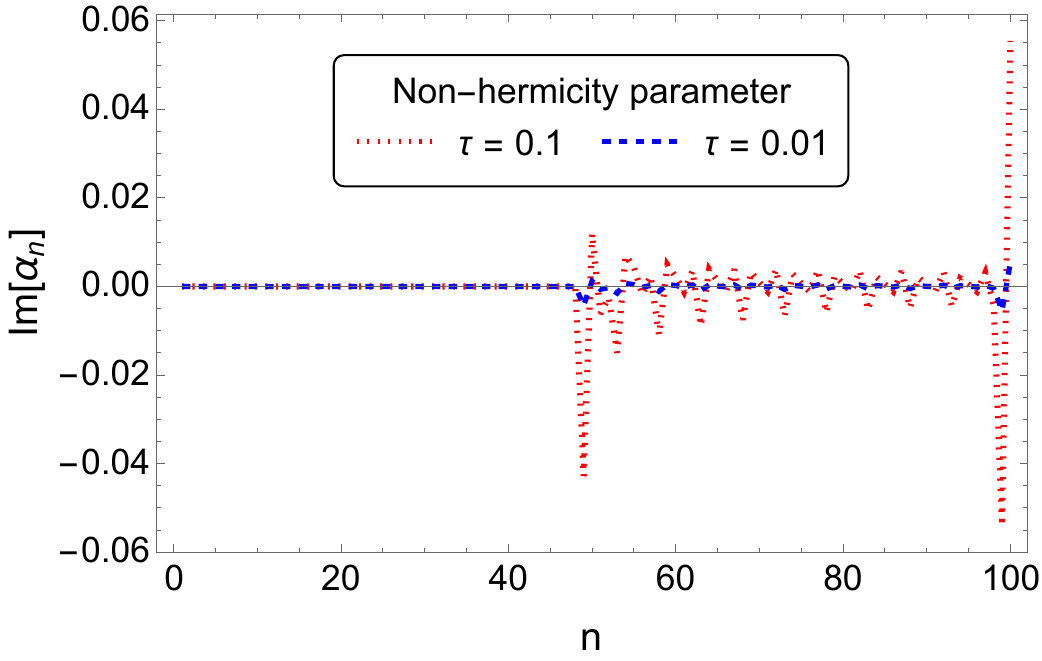}
		\caption{Imaginary parts $\alpha_n$ for $\tau = 0.1$(dotted red) and $\tau = 0.01 $(dashed blue).}
		\label{fig:iman-open-mid}
	\end{subfigure}
	\begin{subfigure}[b]{0.48\textwidth}
		\centering
		\includegraphics[width=\textwidth]{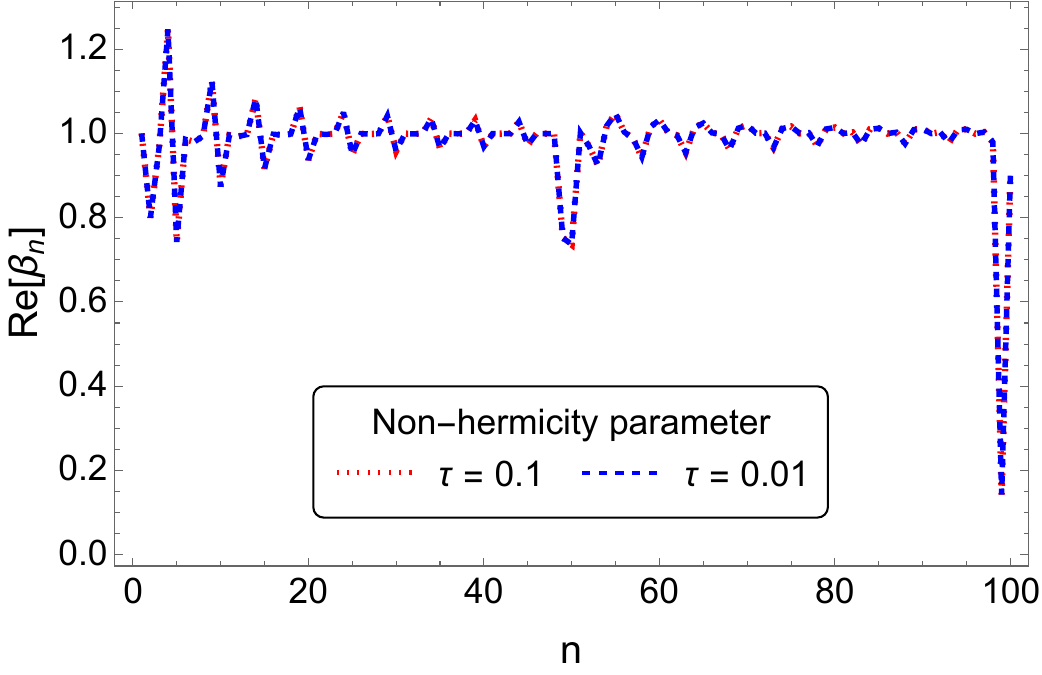}
		\caption{Real parts of $\beta_n$ for $\tau = 0.1$(dotted red) and $\tau = 0.01 $(dashed blue).}
		\label{fig:rebn-open-mid}
	\end{subfigure}
	\hfill
	\begin{subfigure}[b]{0.48\textwidth}
		\centering
		\includegraphics[width=\textwidth]{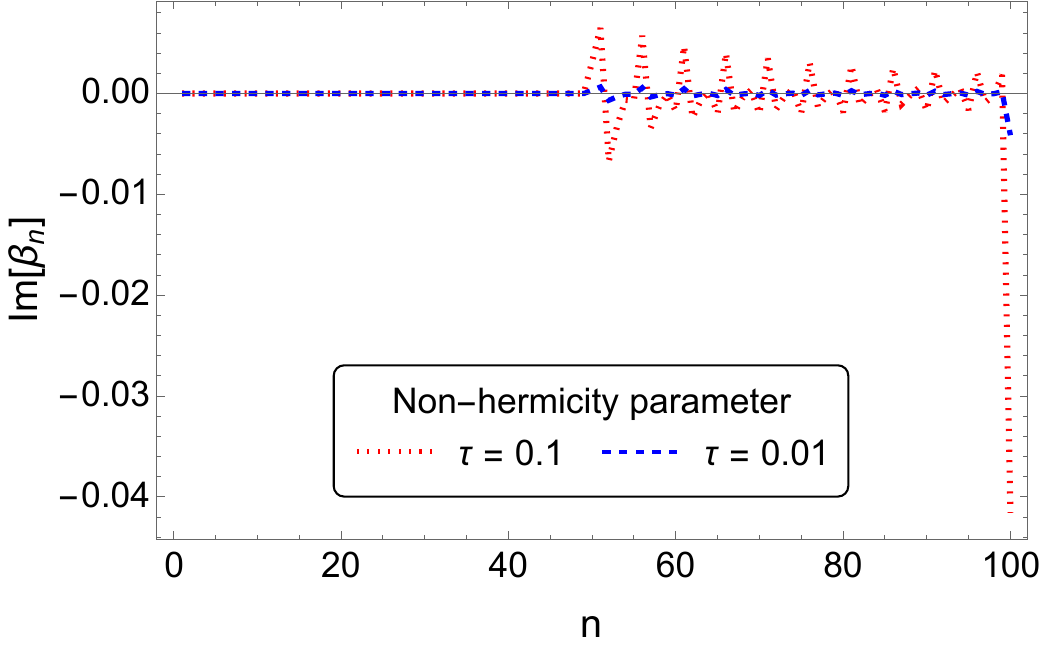}
		\caption{Imaginary parts of $\beta_n$ for $\tau = 0.1$(dotted red) and $\tau = 0.01 $(dashed blue).}
		\label{fig:imbn-open-mid}
	\end{subfigure}
	\caption{Real and imaginary parts of $\alpha_n$ and $\beta_n$ by considering open boundary conditions for different non-hermiticity parameters $\tau$. In these plots, we choose total sites $N= 101$ with a spreading of the initial state between  $48\leq l \leq 52$. The real parts of the coefficients $\alpha_n$ and $\beta_n$ remain invariant under change of $\tau$, while the imaginary parts change showing the effect of the non-hermitian Hamiltonian evolution. }
	\label{fig:anbn-open-mid}
\end{figure}
\begin{figure}[hbtp]
	\centering
	\begin{subfigure}[b]{0.48\textwidth}
		\centering
		\includegraphics[width=\textwidth]{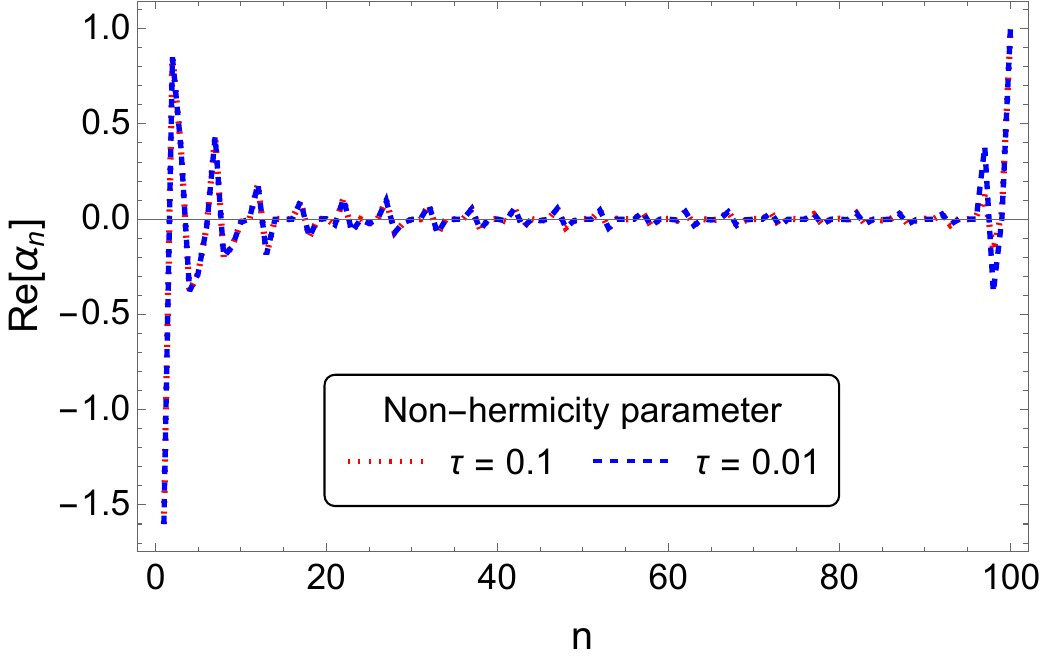}
		\caption{Real parts of $\alpha_n$ for $\tau = 0.1$ (dotted red) and $\tau = 0.01 $ (dashed blue).}
		\label{fig:rean-open-end}
	\end{subfigure}
	\hfill
	\begin{subfigure}[b]{0.48\textwidth}
		\centering
		\includegraphics[width=\textwidth]{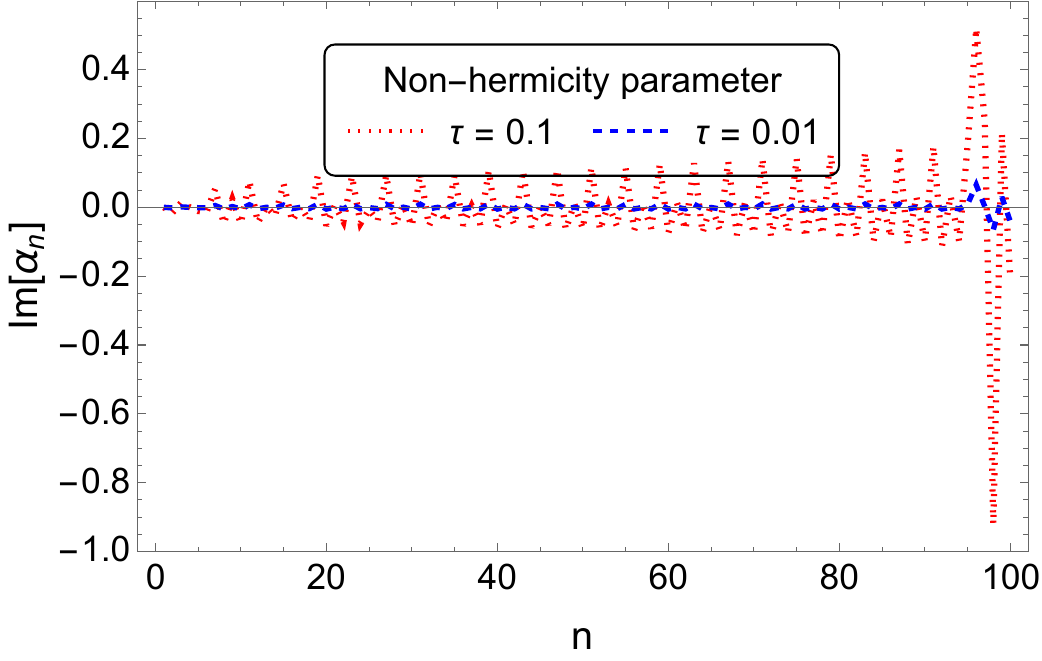}
		\caption{Imaginary parts of $\alpha_n$ for $\tau = 0.1$ (dotted red) and $\tau = 0.01 $ (dashed blue).}
		\label{fig:iman-open-end}
	\end{subfigure}
	\begin{subfigure}[b]{0.48\textwidth}
		\centering
		\includegraphics[width=\textwidth]{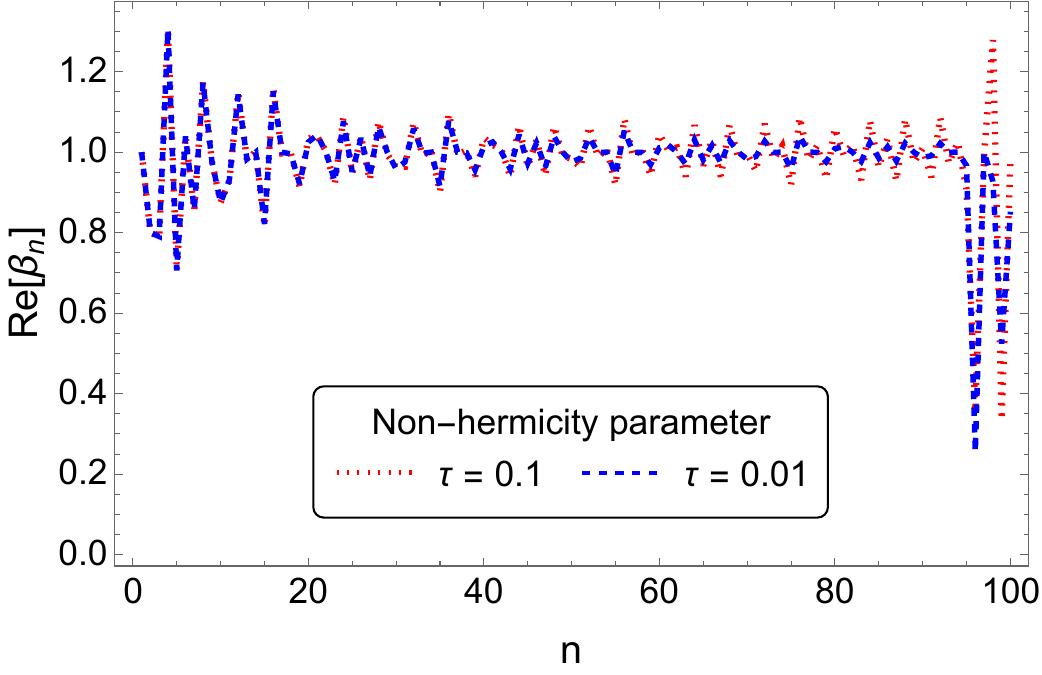}
		\caption{Real parts of $\beta_n$ for $\tau = 0.1$ (dotted red) and $\tau = 0.01 $ (dashed blue).}
		\label{fig:rebn-open-end}
	\end{subfigure}
	\hfill
	\begin{subfigure}[b]{0.48\textwidth}
		\centering
		\includegraphics[width=\textwidth]{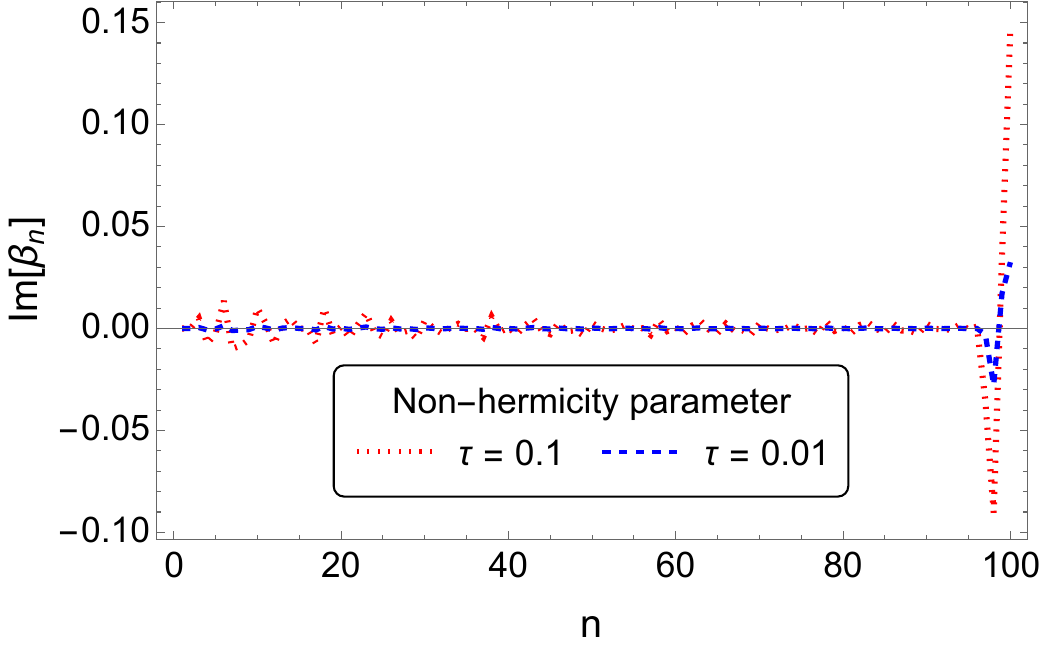}
		\caption{Imaginary parts of $\beta_n$ for $\tau = 0.1$ (dotted red) and $\tau = 0.01 $ (dashed blue).}
		\label{fig:imbn-open-end}
	\end{subfigure}
	\caption{Real and imaginary parts of $\alpha_n$ and $\beta_n$ with open boundary conditions for different values of non-hermiticity parameter $\tau$. Here, the initial state spreading between $95 \leq l\leq 99$ of the total sites $N=101$. For an off-centre spread, the non-hermiticity parameter $\tau$ only substantially changes the imaginary parts of the coefficients.}
	\label{fig:anbn-open-end}
\end{figure}
\begin{figure}[hbtp]
	\centering
	\begin{subfigure}[b]{0.48\textwidth}
		\centering
		\includegraphics[width=\textwidth]{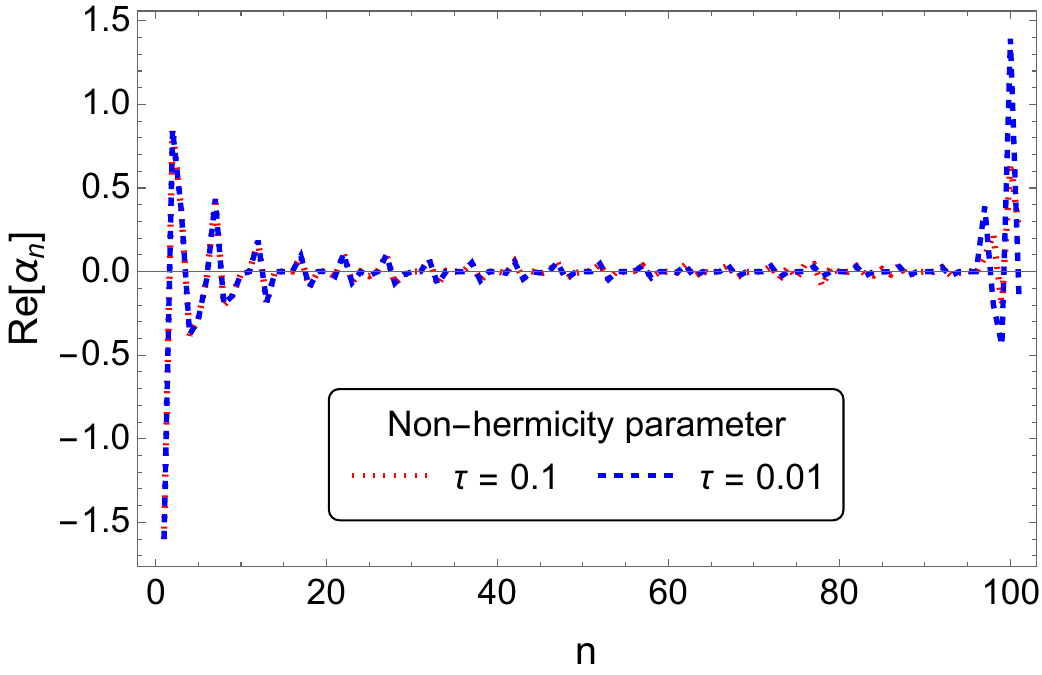}
		\caption{Real parts of $\alpha_n$.}
		\label{fig:rean-closed-ini}
	\end{subfigure}
	\hfill
	\begin{subfigure}[b]{0.48\textwidth}
		\centering
		\includegraphics[width=\textwidth]{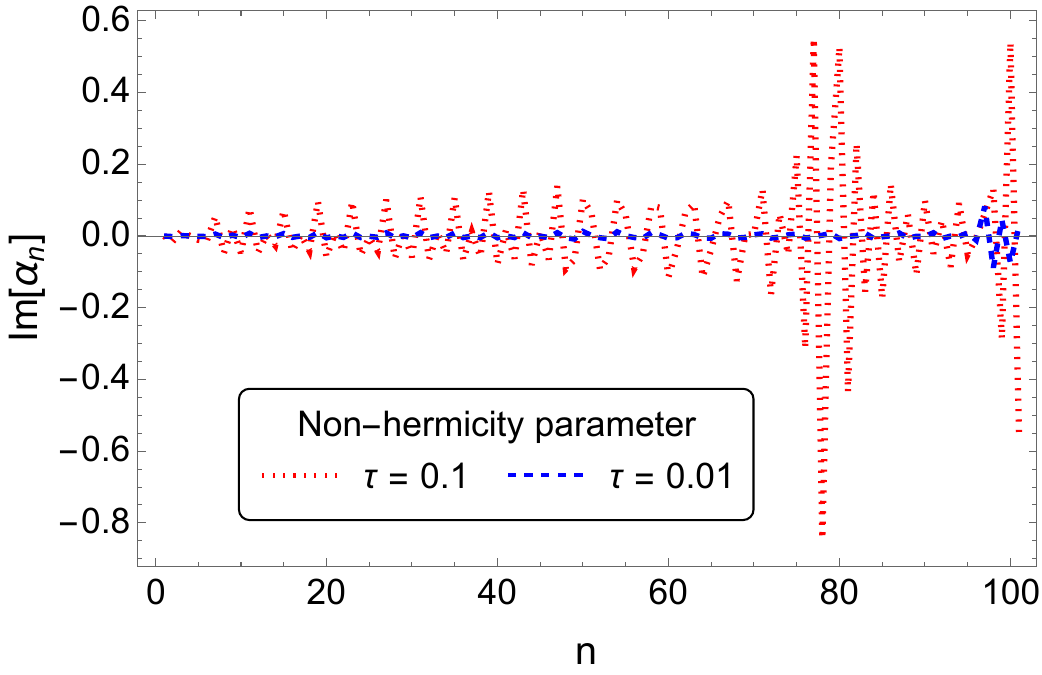}
		\caption{Imaginary parts of $\alpha_n$.}
		\label{fig:iman-closed-ini}
	\end{subfigure}
	\begin{subfigure}[b]{0.48\textwidth}
		\centering
		\includegraphics[width=\textwidth]{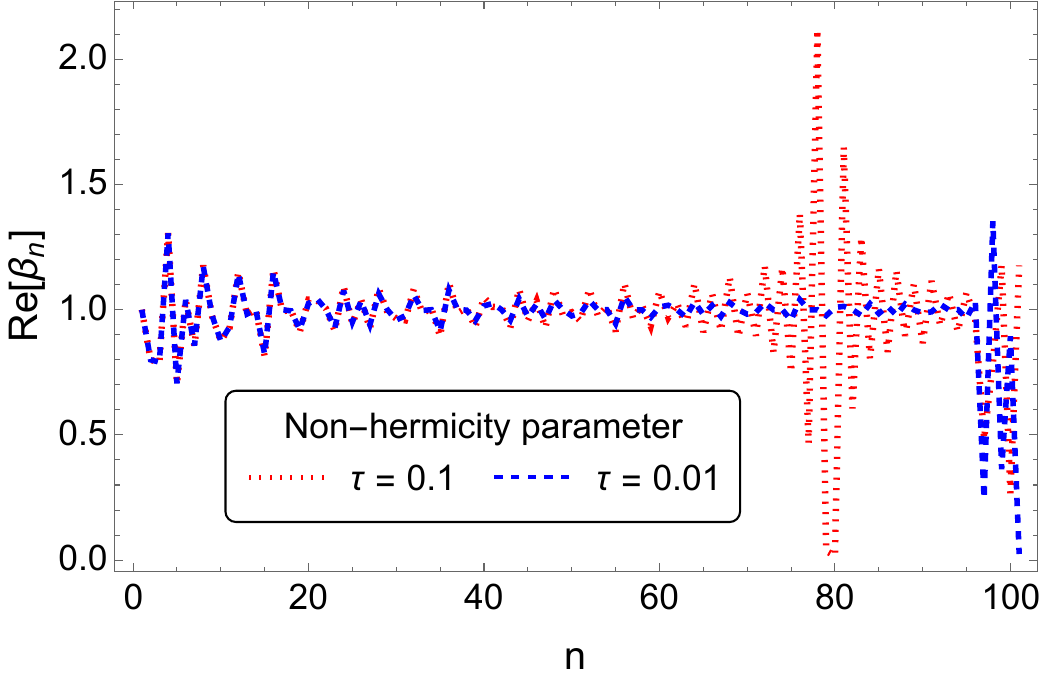}
		\caption{Real parts of $\beta_n$.}
		\label{fig:rebn-closed-ini}
	\end{subfigure}
	\hfill
	\begin{subfigure}[b]{0.48\textwidth}
		\centering
		\includegraphics[width=\textwidth]{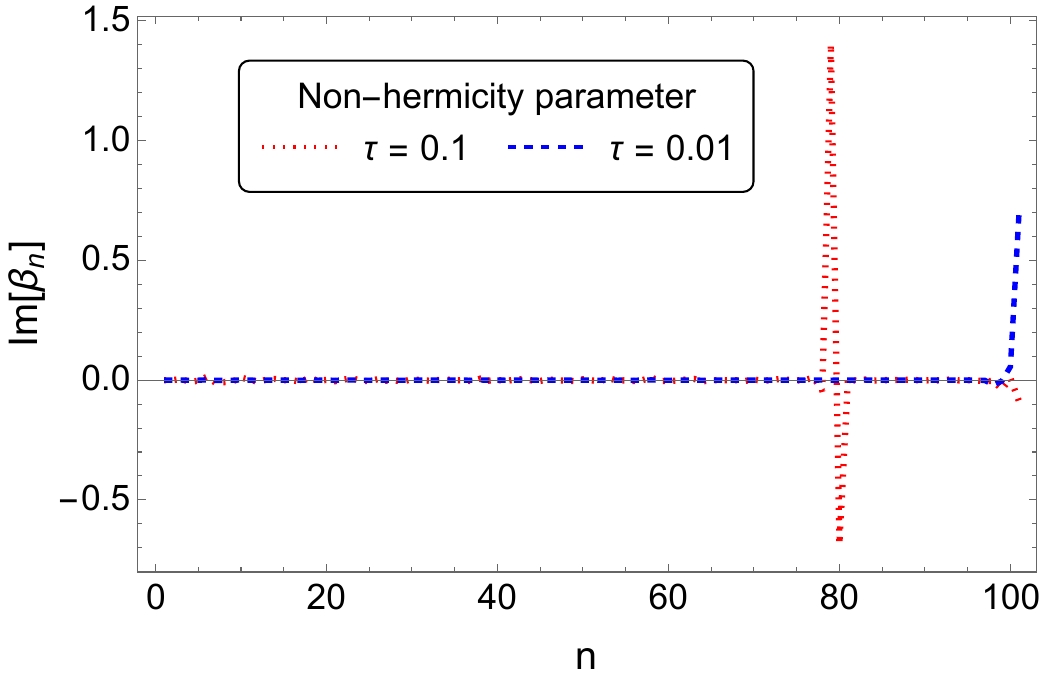}
		\caption{Imaginary parts of $\beta_n$.}
		\label{fig:imbn-closed-ini}
	\end{subfigure}
	\caption{Real and imaginary parts of the $\alpha_n$ and $\beta_n$ for different non-hermiticity parameter $\tau$ values. Here, we choose the number of total sites $N= 102$ with periodic boundary conditions, and initial state localise over $2\leq l\leq 6$. For periodic boundary conditions, the fluctuations in the real part of $\beta_n$ coefficients increase for larger $\tau$ along with the change in the imaginary part for both sets of coefficients. }
	\label{fig:anbn-closed-ini}
\end{figure}
\begin{figure}[hbtp]
	\centering
	\begin{subfigure}[b]{0.4\textwidth}
		\centering
		\includegraphics[width=\textwidth]{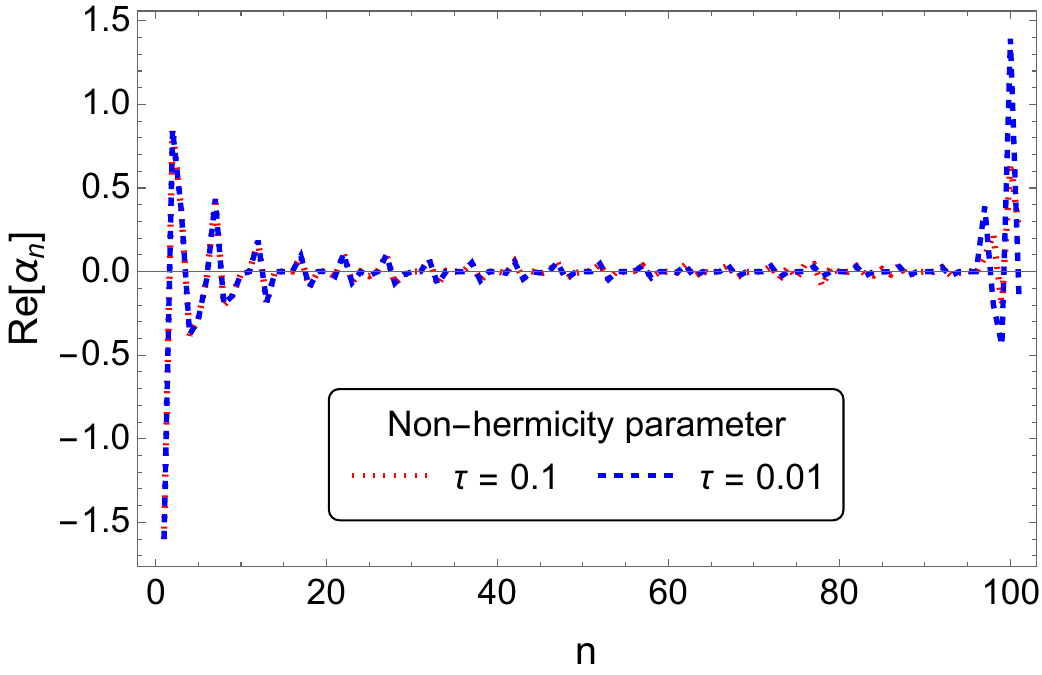}
		\caption{Real $\alpha_n$}
		\label{fig:rean-closed-end}
	\end{subfigure}
	\hfill
	\begin{subfigure}[b]{0.4\textwidth}
		\centering
		\includegraphics[width=\textwidth]{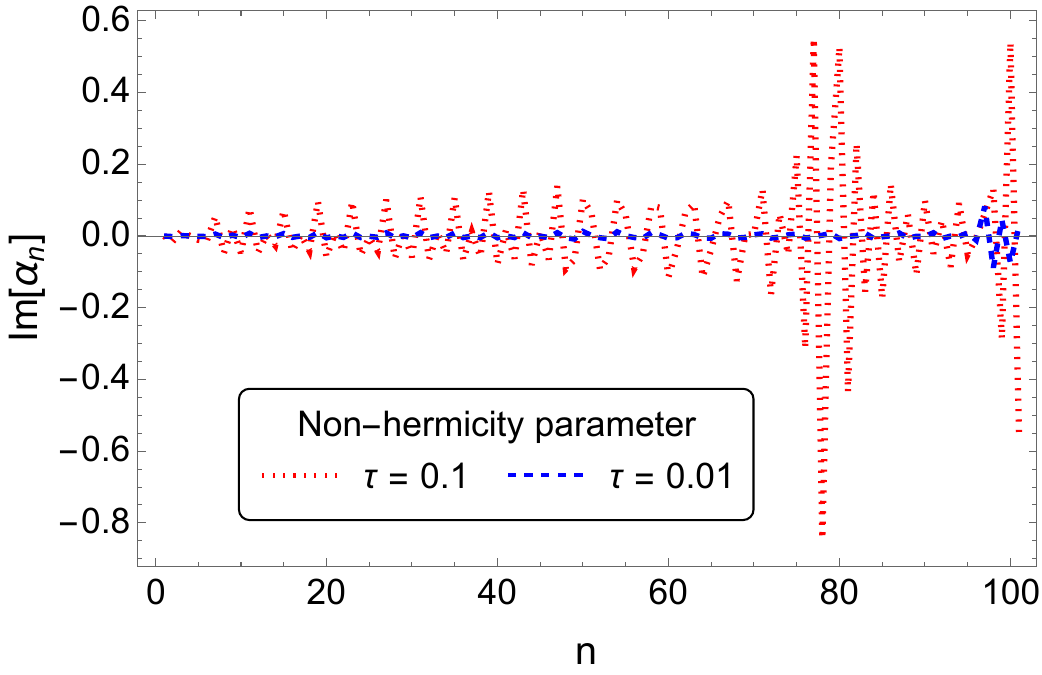}
		\caption{Imaginary $\alpha_n$}
		\label{fig:iman-closed-end}
	\end{subfigure}
	\begin{subfigure}[b]{0.4\textwidth}
		\centering
		\includegraphics[width=\textwidth]{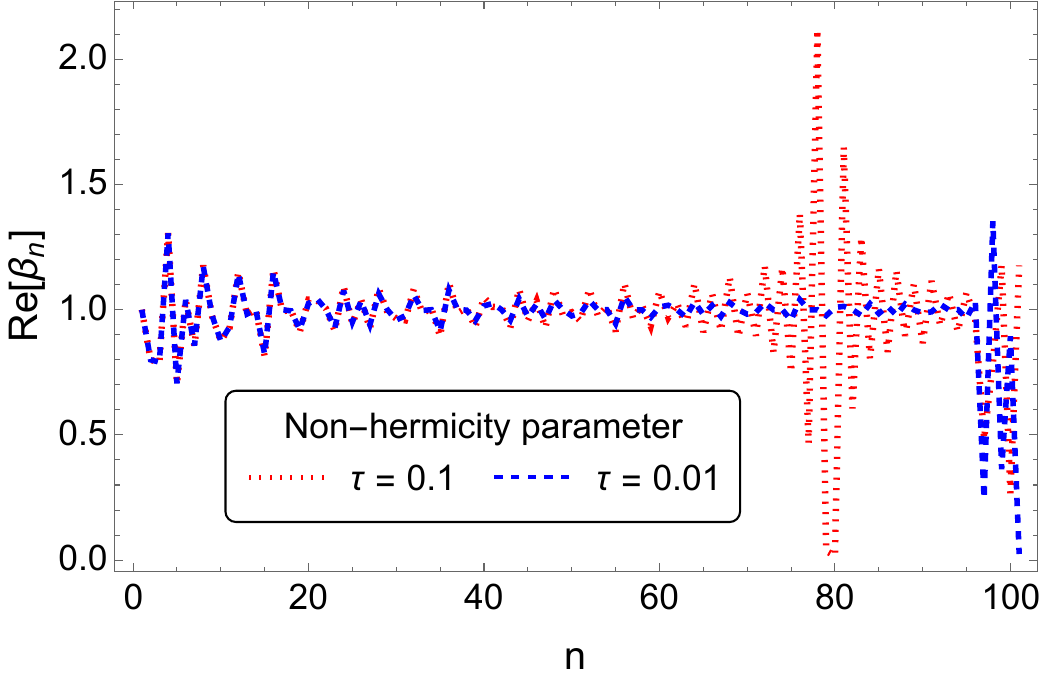}
		\caption{Real $\beta_n$}
		\label{fig:rebn-closed-end}
	\end{subfigure}
	\hfill
	\begin{subfigure}[b]{0.4\textwidth}
		\centering
		\includegraphics[width=\textwidth]{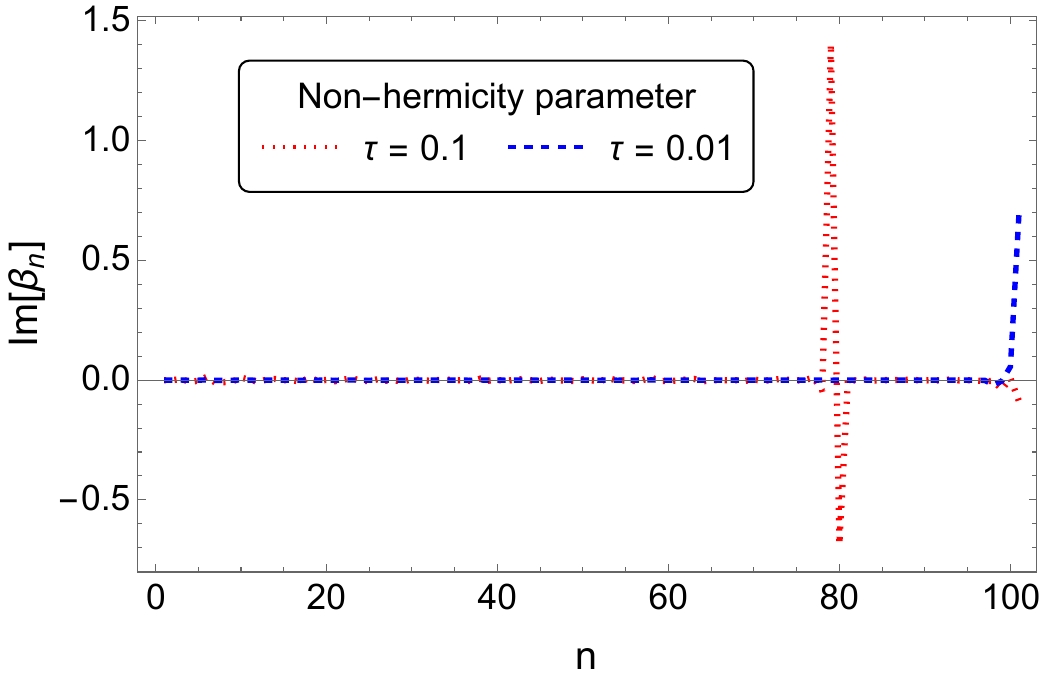}
		\caption{Imaginary $\beta_n$}
		\label{fig:imbn-closed-end}
	\end{subfigure}
	\caption{Real and imaginary parts of the $\alpha_n$ and $\beta_n$ for different non-hermiticity parameter $\tau$ values. Here, we choose the number of total sites  $N = 102$ with periodic boundary conditions, and the spreading of the initial state is $96\leq l\leq 100$. This shows that all real and imaginary Lanczos coefficients are exactly the same as Figure \ref{fig:anbn-closed-ini}. }
	\label{fig:anbn-closed-end}
\end{figure}
In this subsection, we discuss the real and imaginary parts of the Lanczos coefficients $\alpha_n$ and $\beta_n$ for the non-hermitian effective Hamiltonian derived from the complex symmetric Lanczos algorithm. 

We consider open boundary conditions with $N = 101$ total sites while the spreading of initial state is over $48 \leq l \leq 52$ (Figure \ref{fig:anbn-open-mid}) and $95 \leq l \leq 99$ (Figure \ref{fig:anbn-open-end}). We notice that the real parts of the $\beta_n$ and $\alpha_n$ remain nearly unchanged for different degrees of non-hermiticity $(\tau)$. However, the imaginary parts show small deviations. Note that the non-hermitian Hamiltonian  is perturbatively generated with a perturbation parameter. However, from the plots of the probability, spread complexity and entropy in the main text, it is clear that these small deviations give rise to distinguishable features in the above-mentioned quantities. 

We plot the Lanczos coefficients for periodic boundary conditions with total $N= 102$ sites,  and the initial state spread over $2 \leq l \leq 6$ in Figure \ref{fig:anbn-closed-ini}. For the same system size, we also plot the Lanczos coefficients with initial state spread over $96 \leq l \leq 100$ sites in Figure \ref{fig:anbn-closed-end}. From the plots of Lanczos coefficients for the periodic boundary conditions (Figures \ref{fig:anbn-closed-ini} and \ref{fig:anbn-closed-end}), we notice that in terms of the distance from the centre of the chain, the symmetry persists, resulting in exactly identical plots. This symmetry therefore shows up in the spread complexity (Figure \ref{fig:kry_com_diff_spd_per}) and entropy (Figure \ref{fig:kry_ent_diff_spd_per}) plots as well. The origin of this symmetry is the survival amplitude given in Eq.$~$\eqref{survival2}, which is a function of the system size $N$, location of the initial state $l_1\leq l \leq l_2$ and time $t$. The dependence on system size $N$ and the location of the initial spread $l_1,\, l_2$ are such that if two initial states are spread over $l_{11}\leq l \leq l_{12}$, and $l_{21}\leq l \leq l_{22}$ respectively, the survival probability remains same when
\begin{align}
    (l_{22}-l_{21})&=(l_{12}-l_{11}),\\
    \label{symmetrycond}\left|\frac{N}{2}-\frac{l_{22}+l_{21}}{2}\right|&=\left|\frac{N}{2}-\frac{l_{12}+l_{11}}{2}\right|.
\end{align}
The modulus in Eq.~\eqref{symmetrycond} is due to the reason that the spread can be on either side of the centre of the chain. Therefore, the symmetry exists for two initial states, which are of the same width, and the centre of the width has an equal distance from the centre of the chain.

\newpage
\bibliographystyle{JHEP}
\begingroup\raggedright\endgroup

\end{document}